# Background-free measurement of exciton–exciton annihilation by two-quantum fluorescence-detected pump–probe spectroscopy

*Ajay Jayachandran*[1], *Stefan Mueller*[1], *Christoph Lambert*[2], *and Tobias Brixner*[1,3,*]

[1]*Institut für Physikalische und Theoretische Chemie, Universität Würzburg, Am Hubland, 97074 Würzburg, Germany*

[2]*Institut für Organische Chemie, Universität Würzburg, Am Hubland, 97074 Würzburg, Germany*

[3]*Institute for Sustainable Chemistry & Catalysis with Boron (ICB), Universität Würzburg, Am Hubland, 97074 Würzburg, Germany*

AUTHOR INFORMATION

**Corresponding Author**

*E-mail: tobias.brixner@uni-wuerzburg.de

ABSTRACT: We introduce two-quantum (2Q) fluorescence-detected pump–probe (F-PP) spectroscopy as a tool to probe ultrafast multiparticle interactions in many-body systems. We describe a pulse-shaper-based fully collinear setup utilizing phase cycling to capture the 2Q F-PP signal simultaneously with the one-quantum (1Q) F-PP signal. Thus, we investigate the dynamics of energy transfer and diffusion-limited annihilation. We apply a data post-processing strategy to isolate excited-state dynamics from spurious background. The technique is applied to a squaraine heterodimer and a squaraine copolymer to demonstrate the removal of so-called incoherent mixing that generally plagues action-detected nonlinear spectroscopy on multichromophoric systems. Specifically, we show that this approach is not only applicable to 1Q but also to 2Q F-PP signals, eliminating incoherent mixing contributions as well as other “parasitic” signals that result from pulse-overlap ambiguities. As a result, we retrieve background-free spectral and dynamical information of doubly excited electronic states.

## 1. Introduction

Fluorescence-detected pump–probe (F-PP) spectroscopy[1,2] is an action-detected analog to the popular coherently detected transient absorption or pump–probe (PP) spectroscopy[3] for studying dynamics in quantum systems. The F-PP technique circumvents certain limitations encountered in PP experiments. The fluorescence detection allows for target-specific detection of time-resolved nonlinear dynamics of large complex systems with the help of fluorescent probes.[4] The use of transform-limited fs laser pulses in F-PP leads to a high temporal resolution while retaining a large spectral bandwidth at the same time, enabling one to acquire ultrafast dynamical processes such as exciton transfer and annihilation before spontaneous emission. This poses a crucial advantage compared to other established methods such as fluorescence upconversion, where the temporal resolution is limited by the phase-matching bandwidth of a nonlinear crystal. The data collected in F-PP is free from non-resonant solvent response which otherwise complicates the interpretation of early-time dynamics in PP measurements,[5] making F-PP suitable for studying fast kinetic processes including those associated with many-body correlations. The signatures of these correlations can generally be inferred from the higher-order nonlinear response of the system.[6,7] Probing the higher-order response provides direct access to quantities such as energies, transition-dipole moments and linewidths of higher-excited states, as well as coupling strengths and energy-transfer pathways between them. The higher-order response also inherently contains the information obtainable from lower-order methods. The incoherent detection in F-PP yields a high sensitivity for monitoring ultrafast processes for microscopy experiments,[8] even down to the single-molecule level,[9] because the absence of a strong probe background makes the detection of weak signals easier than with transient absorption.[10]

Coherently detected PP experiments involve a single delay scan between a strong "pump" pulse whose electric field interacts an even number of times with the sample and a weak "probe" pulse that interacts once with the system under study in a perturbative description of light–matter interaction. The interpulse delay $T$ tracks the evolution of the system population following pump excitation. The emitted signal is heterodyne detected, i.e., the probe itself acts as a local oscillator after the final field interaction and is then spectrally dispersed to produce a frequency axis that we will refer to as $\omega_t$. The interpretation of ultrafast PP and its frequency-resolved extension, coherent two-dimensional spectroscopy,[11–14] benefits from the theoretical framework for nonlinear optical spectroscopy provided by Shaul Mukamel and Minhaeng Cho.[15–26,12,27,27,28] Building on these concepts and insights on how spectroscopic methods are understood and applied, we revisit PP in light of a fluorescence-detected implementation.

In contrast to coherently detected PP, F-PP uses one pump pulse and two probe pulses, resulting in a three-pulse experiment. This excitation sequence may bring the system into an excited state from which spontaneous fluorescence can be emitted. This fluorescence encodes the nonlinear response, which in turn contains information on the dynamics evolving during all the time delays. Here, the pump–probe delay between the pump and the first probe pulse, $T$, is also referred to as the population delay, and the delay between the two pulses of the probe pair, $t$, provides the $\omega_t$ detection frequency axis after Fourier transformation. It is also possible to conduct a two-pulse experiment with fluorescence detection to measure quantum beats as demonstrated by Bruder et al. on the example of diluted gas-phase systems.[29,30] Though this method is highly sensitive to coherent oscillations, the use of one probe pulse omits frequency resolution for the probe step and pure transients of population decay cannot be received. To obtain the pure transient signal in fluorescence-detected pump–probe (F-PP) measurements, one begins with the fluorescence intensity recorded when all three pulses are present. From this, two components are subtracted: (1) the fluorescence induced by the two probe pulses alone, obtained by chopping the pump pulse, and (2) the pump-only-induced fluorescence, estimated as the average fluorescence intensity across the $t$ delay interferogram at each $T$ delay step. This separation works because the F-PP signal exhibits oscillations as a function of $t$, whereas the pump-only-induced fluorescence remains constant with respect to $t$. The oscillatory component of the signal (including any positive offset from the probe-pair-only contribution) is first isolated by measuring the fluorescence with the pump chopped and subtracting this probe-only background. As a result, averaging over $t$ effectively isolates and removes only the stationary pump-only background, leaving the desired oscillatory F-PP response intact. F-PP spectra offer the advantage that they are practically free of excited-state absorption (ESA) contributions and thus only consist of ground-state bleach (GSB) and stimulated emission (SE).[1,31] Because ESA enters the total signal with a different sign, the overlap of ESA with GSB and SE contributions can complicate the analysis of C-PP spectra. The advantage of ESA-free spectra is particularly pronounced in multichromophoric systems which feature a rich excited-state landscape and thus a multitude of ESA pathways.

In fluorescence-detected two-dimensional electronic spectroscopy (F-2DES), an additional pump pulse is used to create the excitation axis of a 2D spectrum by Fourier transformation with respect to the time delay between the two pump pulses.[32–35] In comparison, F-PP omits the frequency resolution for the excitation step but retains the frequency resolution for the detection. While F-PP thus provides less information compared to F-2DES in terms of the homogenous linewidth analysis of excitations, it can still be useful because the measurement time is reduced, and in some cases, one integrates over the 2D excitation axis anyway for analysis. For example, Sahu et al.[2] have experimentally shown how the vibrational dynamics probed over the *T* delay in the F-2DES spectrum can be extracted with an F-PP experiment created by keeping the pump pair delay of the F-2DES pulse sequence to zero, hence reducing the total data acquisition time.

Prior F-PP experiments were performed in a partially collinear geometry comprising a collinear probe pair and a noncollinear pump[1] as well as in a fully collinear geometry using a combination of a common-path interferometer with a delay stage[31] or by using two nested Mach–Zehnder interferometers.[2,8] In those projects, the focus was on the extraction of the fourth-order nonlinear signal, in which a one-quantum (1Q) coherence is probed during *t* to describe the single-excitation dynamics at both ensemble and single-molecule levels.

It is also possible to probe multiple-quantum coherences simultaneously with 1Q coherence in F-PP.[6,7,36] Two-quantum (2Q) coherences oscillate at around twice the center frequency of the excitation pulse.[37–41] The 2Q signal probed in F-PP provides crucial insights into the energetics, coupling, and dynamics of doubly excited states.[6,29,30,36,42] The signal requires a total of six or more, even-valued, number of field interactions in total from pump and probe pulses together. The 2Q signal previously described in the light of F-2DES has been identified to be useful in studying multiparticle interactions in many-body systems. The signal was used to extract the dynamics of exciton transport via direct probing of exciton–exciton annihilation.[6,7] This requires at least a sixth-order nonlinear process, i.e., four interactions to generate and two interactions to probe a doubly excited population.

In this work, we introduce a new method for extracting both 1Q (Section 2.1) and 2Q (Section 2.2) signals in F-PP, using a fully collinear excitation and a pulse shaper. In this geometry, the extraction of nonlinear signals requires a systematic incrementation of the pulse phases and subsequent analysis of linear combinations of raw data, also known as phase cycling, that can be carried out in a nested fashion[43] or via so-called cogwheel phase cycling as we demonstrated recently.[7,44] The 1Q F-PP signal provides access to the single-excitation dynamics on a fast time scale (< 100 fs). The 2Q F-PP signal simultaneously provides information on the correlations between pairs of such excitations. We also discuss the origin of undesirable static and dynamic backgrounds that plague the study of multiparticle dynamics in action-detected spectroscopy. The static background, also called "incoherent mixing,"[45,46] generally limits the retrieval of excited-state dynamics information, while the dynamic background that is associated with parasitic multi-quantum coherences[47,48] generally complicates the analysis of multiparticle correlations. We describe these contributions in the context of 2Q F-PP and show how our method overcomes the limitations. For that purpose, we discuss general properties of the 1Q and 2Q F-PP signals with aid of nonperturbative simulations (Section 3.1) and experimentally demonstrate the simultaneous acquisition of both signals on the example of a well-studied squaraine dimer and a related squaraine copolymer (Section 3.2). As the copolymer represents a multichromophoric system which is subject to a somewhat larger extent of incoherent mixing, we particularly illustrate the removal of such undesired contributions using our approach.

## 2. Method development

### 2.1. One-quantum fluorescence-detected pump–probe spectroscopy

Using phase cycling and extracting the dominating fourth-order F-PP signal,[1,8] one can probe singly excited and ground-state population dynamics along the $T$ delay while 1Q coherences are probed along the $t$ delay (Figure 1a, top). Since 1Q coherences are probed, we refer to this as the "1Q F-PP" signal. In phase cycling, each nonlinear signal is assigned a distinct phase signature, defined as a specific linear combination of the absolute phases of the pulses in a three-pulse sequence including one pump pulse ($\varphi_{\text{pump}}$) and two probe pulses ($\varphi_{\text{probe1}}$ and $\varphi_{\text{probe2}}$). These phase combinations act as "tags" that allow different signal pathways to be selectively extracted from the total measured signal. To isolate the desired 1Q F-PP fourth-order signal, one detects the response with the phase combination "$+\varphi_{\text{pump}} - \varphi_{\text{pump}} + \varphi_{\text{probe1}} - \varphi_{\text{probe2}}$". Note that in this shorthand notation, the first two terms with opposite sign do not just cancel each other, but in fact they represent individual electric field interactions of the pump pulse with the sample in a perturbative description of fourth order. This desired signal can be obscured, however, by the overlapping second-order signal "$+\varphi_{\text{probe1}} - \varphi_{\text{probe2}}$" that really only results from interactions with the probe pulses.[31] To separate these contributions, the raw data is collected using a sequence of measurements in which the phases of the pump and probe pulses are systematically varied across multiple phase steps. Specifically, the pump pulse is chopped on and off in alternating measurements, allowing for the subtraction of the second-order background. By calculating linear combinations of these measurements according to the known phase signatures, one can isolate and reconstruct the pure fourth-order nonlinear signal. The 1Q F-PP experiment using phase cycling also ensures that signals which do not contain these specific phase dependencies such as pump-only-induced fluorescence contributions are inherently suppressed. The pump-only signal originates from a single pulse and therefore does not exhibit any phase dependence involving probe pulses in the pulse sequence. This is different from other approaches to measure 1Q F-PP,[1] that require this component to be explicitly removed during the subtraction.

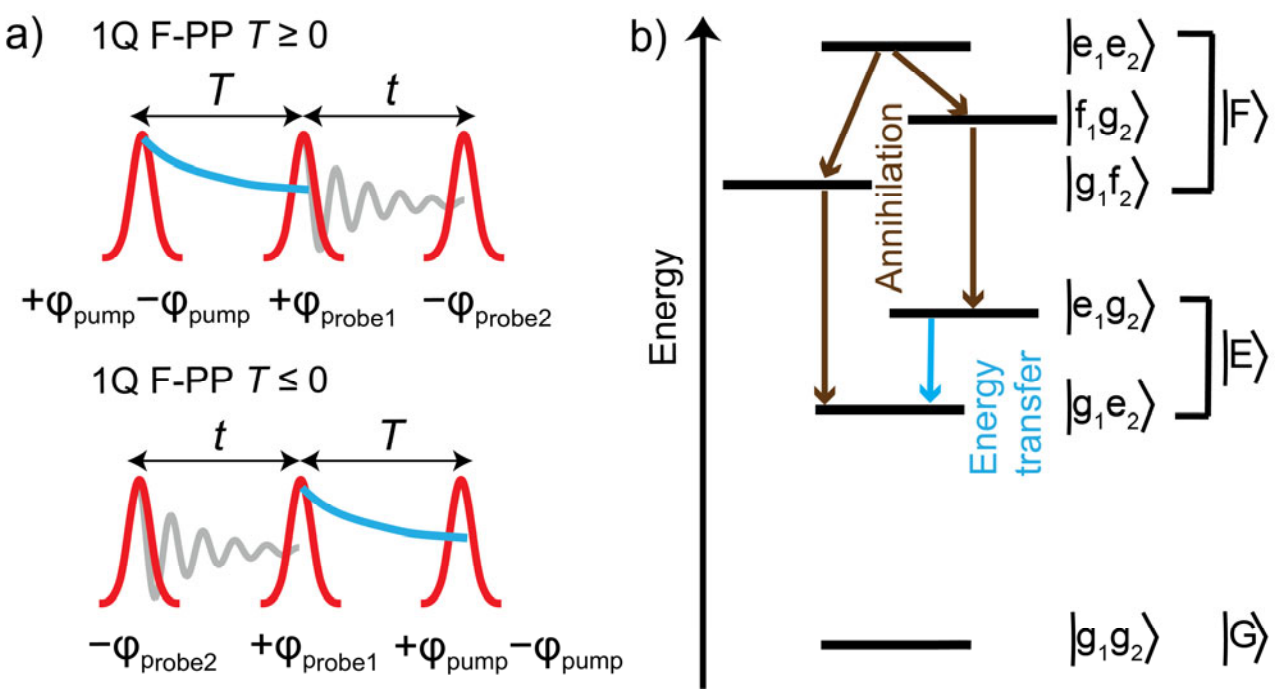

**Figure 1.** One-quantum (1Q) F-PP spectroscopy and exemplary quantum system. (a) Pulse sequences with unique phase signatures described by the phase imparted from individual pump-pulse ($\varphi_{pump}$) and probe-pulse interactions ($\varphi_{probe1}$ and $\varphi_{probe2}$) for capturing the 1Q F-PP signal. A train of three pulses (red) is shown with mutual delays $T$ and $t$ and time ordering between pump and probe-pulse pairs for $T \geq 0$ (top) and $T \leq 0$ (bottom). We have 1Q coherences (gray) that evolve during $t$ for 1Q F-PP. During $T$, population dynamics evolve between various excited states (light blue). (b) Energy level manifold of an exemplary dimer system. Annihilation (brown arrows) occurs from the doubly excited ($|F\rangle$) manifold with individual states $|e_1e_2\rangle$, $|g_1f_2\rangle$, and $|f_1g_2\rangle$ to the singly excited manifold ($|E\rangle$). Energy transfer (light blue arrow) occurs between singly excited states $|e_1g_2\rangle$ and $|g_1e_2\rangle$. The ground state ($|G\rangle$) is represented by $|g_1g_2\rangle$.

In coherently detected PP spectroscopy, a reversal of the time ordering, i.e., the probe pulse interacting before the pump pulse as denoted by a negative $T$ delay, yields the perturbed free-induction decay.[49,50] Likewise in 1Q F-PP, when the probe pair interacts before the pump pulse (Fig. 1a, bottom), the resulting transient spectra also hold important information. The signal at negative $T$ is considered as fluorescence excitation spectrum because the probe pulse, arriving before the pump, is solely responsible for exciting fluorescence.[1] The delayed pump then interacts with the already-excited population and reduces the fluorescence. By measuring how much the pump quenches the fluorescence at different probe wavelengths, the technique effectively maps how efficiently each probe wavelength excites the system, just as in a conventional fluorescence excitation spectrum. The measurement of the 1Q F-PP spectrum at both negative and positive $T$ delays thus can provide detailed spectroscopic insights into a system.[1,51,31]

To analyze all relevant contributions to the fourth-order 1Q F-PP nonlinear response in the density-matrix formulation, we first consider for illustration an exemplary many-body system that consists only of two subsystems 1 and 2, such as a molecular dimer. Both subsystems can undergo excitation upon illumination and individually each contain a ground state, $|g_i\rangle$, a singly excited state, $|e_i\rangle$, and a doubly excited state, $|f_i\rangle$, where the subscript labels the subsystem, $i$ = 1,2. The combined system (Figure 1b) then contains three manifolds of electronic states consisting of a common ground state $|G\rangle$ where both subsystems are in the ground state, $|g_1g_2\rangle$, a manifold $|E\rangle$ of two singly excited states where either one of the subsystems is in the first excited state, $|g_1e_2\rangle$ and $|e_1g_2\rangle$, and a manifold $|F\rangle$ of doubly excited states, $|e_1e_2\rangle$, $|g_1f_2\rangle$, and $|f_1g_2\rangle$, with fast relaxation of the doubly excited manifold to the singly excited manifold. We do not consider triply or higher excited states of the combined system here, because their population would require triple excitation and the potential pathways do not contribute to the signals investigated here, at our choice of pump and probe intensities, as explained in more detail in Section 2.2. The $|g_1f_2\rangle$ and $|f_1g_2\rangle$ states signify that either of the two subsystems is doubly excited while the other is in the ground state, whereas the $|e_1e_2\rangle$ state involves a single excitation of both subsystems. Note that we do not take into account explicitly the energy shifts associated with electronic coupling between subsystems and the associated formation of exciton states upon diagonalization of the total Hamiltonian. The level scheme and Feynman diagrams to be discussed below are just meant to illustrate and categorize the evolution of coherences and populations in the product-state density matrix of the system as a result of the applied pulse sequences. For a full picture, the labeled product states can be used to construct exciton states within a suitable linear combination.

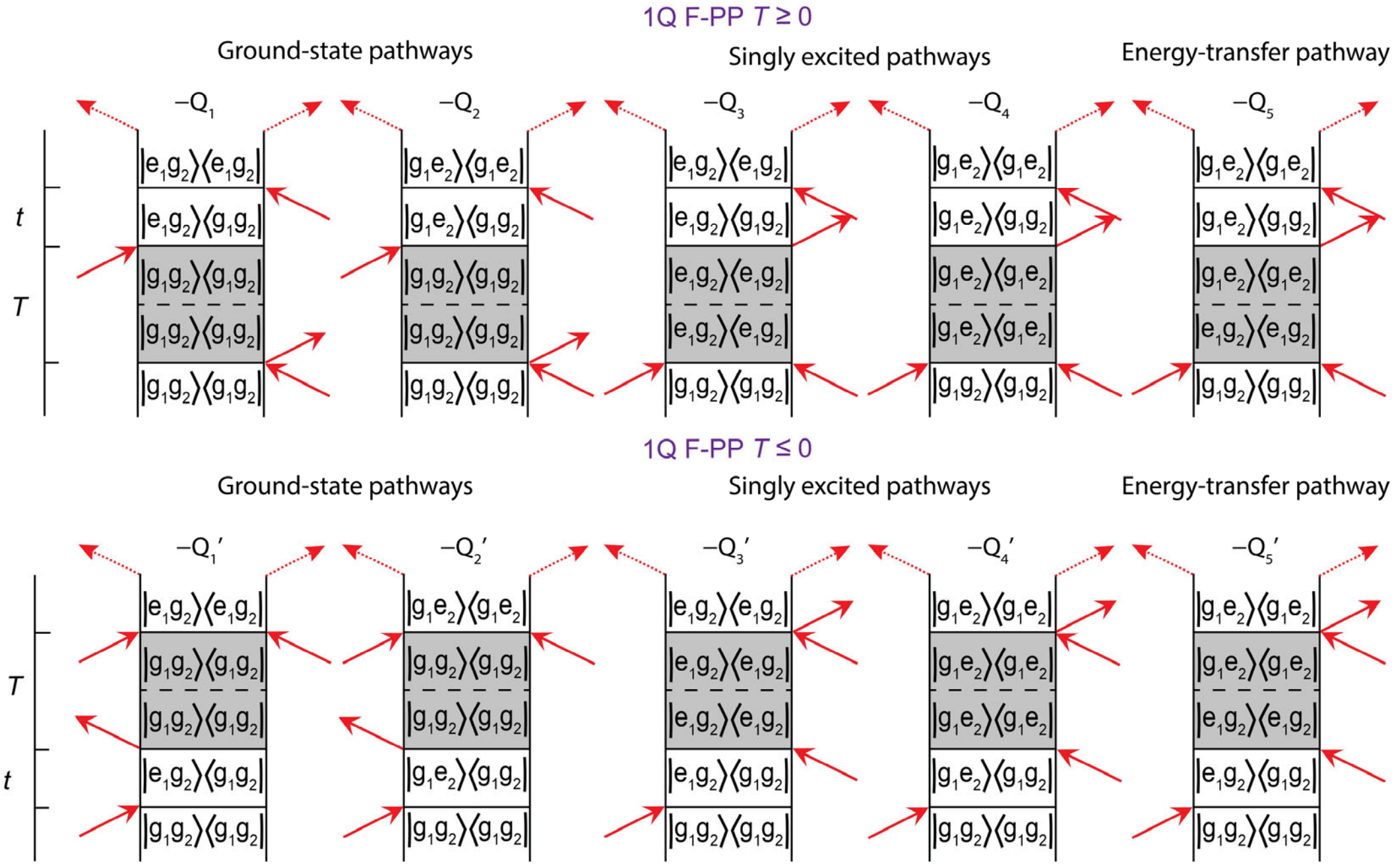

**Figure 2.** Double-sided Feynman diagrams for 1Q F-PP in the case of a dimer. Top: Feynman diagrams for positive pump–probe delay assuming fast relaxation from the doubly excited manifold to the singly excited manifold, thus leading to the cancellation of excited-state absorption pathways. The dark-shaded region in each diagram highlights population dynamics. Bottom: 1Q F-PP Feynman diagrams for negative pump–probe delay. For ground-state pathways ($Q_1$, $Q_2$, $Q_1$’, $Q_2$’) and singly excited pathways ($Q_3$, $Q_4$, $Q_3$’, $Q_4$’), the population of the system evolves in the ground and the singly excited state, respectively. All pathways except for the energy-transfer pathways $Q_5$ and $Q_5$’ cancel off for the difference spectrum Δ1Q F-PP ($\omega_t$, $T$).

We depict the 1Q F-PP excitation pathways through the system in Liouville space in Figure 2 using double-sided Feynman diagrams, under the assumption of the rotating-wave approximation[52]. The 1Q F-PP diagrams for $T \geq 0$ (Figure 2, top row) and $T \leq 0$ (Figure 2, bottom row) contain pairs of pathways with identical nonlinear response. Specifically, this applies to the ground-state pathways ($Q_1$ and $Q_1$’ as well as $Q_2$ and $Q_2$’) and the singly excited pathways ($Q_3$ and $Q_3$’ as well as $Q_4$ and $Q_4$’). In these pathway pairs, the population evolves in the ground state or the same singly excited state during $T$. Each member of such a pathway pair is expected to occupy the same spectral position along the Fourier-transformed $\omega_t$ detection axis as its associated twin, which in turn leads to their mirror-symmetric distribution relative to $T = 0$ in the 1Q F-PP spectrum.[51] Note that there are additional pairs of singly excited pathways for both $T \geq 0$ and $T \leq 0$ which are not shown in Fig. 2, however, these pathway pairs cancel each other off due to their opposite signs. This cancellation is a result of the assumption of rapid relaxation from the doubly to the singly excited manifold.

In contrast, the $Q_5$ and $Q_5'$ pathways are not identical in the two halves of the *T* delay. These pathways describe the resonant energy-transfer process from the $|e_1g_2\rangle$ to the $|g_1e_2\rangle$ excited state. While the 1Q coherence in $Q_5'$ oscillates with the transition frequency from $|g_1g_2\rangle$ to $|e_1g_2\rangle$, i.e., the energetically higher-lying of the two $|E\rangle$ states (Figure 1b), the coherence in $Q_5$ oscillates with a somewhat lower frequency corresponding to the transition from $|g_1g_2\rangle$ to $|g_1e_2\rangle$. For finite electronic coupling between the two subsystems, the product states $|g_1e_2\rangle$ and $|e_1g_2\rangle$ combine to form excitonic states $|A\rangle$ and $|B\rangle$. In a dimeric system, the 1Q F-PP transient map then contains spectral features corresponding to energy transfer between these excitonic states located at energies of these states along the detection axis.

Utilizing the symmetry in the distribution of ground-state pathways and singly excited pathways across negative and positive *T* delays for extracting pure excited-state dynamics in 1Q F-PP was discussed previously.[51] In the particular context of a dimer, this symmetry can be exploited to extract the energy-transfer dynamics between the singly excited states by constructing a difference map and exploiting the temporal symmetry with respect to $T = 0$. To do that, we record F-PP measurements as two-dimensional maps, with *T* represented on the horizontal axis and the $\omega_t$ detection frequency on the vertical axis. The full dataset includes both negative and positive values of *T*, symmetrically distributed around zero. We then construct a difference map by reflecting the negative-time portion of the dataset across $T = 0$ and subtracting it from the corresponding positive-time data. Specifically, for each fixed value of $\omega_t$, the signal at negative time $-T$ is mapped onto the corresponding signal for positive time $T$ and subtracted. The resulting difference signal, Δ1Q F-PP ($\omega_t$, $T$), defined for $T \geq 0$, is given by:

$$\Delta 1\text{Q F-PP } (\omega_t, T) = 1\text{Q F-PP } (\omega_t, T) - 1\text{Q F-PP } (\omega_t, -T). \quad (1)$$

While the subtraction will remove the contributions of ground-state pathways and singly excited pathways due to their symmetrical distribution with respect to $T = 0$, the energy-transfer pathways are asymmetrically distributed and will remain after subtraction. This circumstance allows us to selectively extract the energy-transfer dynamics from Δ1Q F-PP ($\omega_t$, $T$) by fitting the data as a function of $T$ for a particular, fixed, $\omega_t$ at a spectral position corresponding to, e.g., an excitonic peak. It is important to note that the validity of the subtraction in Eq. (1) to remove ground-state pathways and singly excited pathway contributions is subject to the assumption that the system temporally evolves without any memory of the bath interactions, in order to retain the symmetry of these contributions across $T \geq 0$ and $T \leq 0$.[6] Charvátová et al.[51] also recently described this energy-transfer extraction procedure for 1Q F-PP, along with a mathematical derivation of the underlying subtraction. They also discussed F-2DES, for which they demonstrated that integrating a F-2DES spectrum along the excitation axis yields the 1Q F-PP ($\omega_t$, $T$) spectrum, while integrating along the detection axis yields the equivalent 1Q F-PP ($\omega_t$, $-T$) spectrum.

The description of the single-particle dynamics captured in 1Q F-PP can be generalized from a dimer as in Figures 1b and 2 to a system containing $N$ (coupled) subsystems. For that purpose, we broadly classify the Feynman pathways into self- and cross-population pathways. In a 1Q F-PP experiment, the incoherent signal used to decode the nonlinear response is detected by integrating the photon counts over the entire temporal window of fluorescence emission. The self-population pathways represent the interaction of all laser pulses with the same subsystem. In self-population pathways, a fourth-order excited-state population is created that ideally does not undergo any further dynamics during the temporal window of fluorescence detection, aside from nonradiative relaxation followed by radiative decay. These pathways are further subdivided into self-population ground-state pathways ($GSB^s$) and singly excited pathways ($SE^s$) based on whether the system evolves across ground or singly excited states during $T$. The cross-population ground-state pathways ($GSB^c$), as described by Bruschi et al.,[46] involve excitation events occurring at two distinct subsystems. In F-PP with one pump and two probe pulses, two interactions of the pump pulse may occur at one subsystem, while the subsequent interactions from the probe pulses occur at a different subsystem. Among a population of $N$ subsystems, this means that the excitation and probing processes can occur across different subsystems, allowing for cross-population pathways.

The $GSB^c$ pathways represent the incoherent mixing[53] of linear signals emitted during the temporal window of fluorescence detection. This mixed signal is not produced by the fourth-order population but rather reflects nonlinear population dynamics produced by linear light–matter interactions on two subsystems of a weakly coupled assembly. The incoherently mixed signals are modulated with the same frequency and phase as the self-population $GSB^s$ and $SE^s$ fourth-order response. Hence, the $GSB^c$ pathways cannot be separated from the $GSB^s$ and $SE^s$ pathways of the nonlinear response by using phase cycling, as they both carry the same phase signature. While the singly excited and energy-transfer pathway contributions are weighted each by a factor of $N + (N - 1)$ for a system with $N$ subsystems, the contribution of ground-state pathways to the overall signal response scales with a factor of $N(N - 1) + N$.[45] In the case of systems with large $N$ and fast annihilation, the $GSB^c$ pathways thus obscure the retrieval of excited-state dynamics in fourth-order F-PP spectroscopy, as they are indistinguishable and inseparable from regular $GSB^s$ pathways. In particular, the ratio of the total number of ground-state pathways ($GSB^s + GSB^c$) to excited-state pathways (i.e., singly excited pathways plus energy-transfer pathways) for the system scales linearly with a factor of $\approx N$.[45] This circumstance was identified as a complication for studying excited-state dynamics using any action-detected spectroscopy of systems with a large number of subsystems. This is because the additional $GSB^c$ pathways do not help in probing the single-excited-state population dynamics during $T$ but dominate the overall signal intensity and remain as a constant background.[54] Note that the classification of the ground-state pathways as $GSB^s$ and $GSB^c$ is only applicable in the limit of weakly interacting subsystems. This is because in the presence of complete delocalization, the system would act as one whole absorbing unit and, therefore, can be represented by the same number of ground-state and singly excited pathways. Hence, for an arbitrary system, the ratio of the total number of ground-state pathways to singly excited pathways is equal to the number of subsystems $N$ over which excitations are delocalized.[54] This quantification has been identified as a useful feature in action-detected experiments compared to coherently detected spectroscopy.[46,54,55]

### 2.2. Two-quantum fluorescence-detected pump–probe spectroscopy

The use of phase cycling in F-PP also allows us to extract higher-order signals with 2Q coherences, i.e., coherences between the ground state and a doubly excited state, that evolve along $t$, which goes beyond previous work. Such signals cannot be obtained in coherently detected PP spectroscopy. Although doubly excited states can be accessed via two-photon interactions, they usually undergo nonradiative relaxation due to fast internal conversion to the singly excited state following Kasha's rule. As heterodyne detection requires radiative transitions to generate a coherent signal field,[56,57] spectral signatures of 2Q coherences involving doubly excited states cannot be observed along the probe axis in coherently detected PP spectroscopy. These coherences can rather be observed as part of the coherence spike which occurs during pulse overlap, where they manifest themselves as oscillations of the PP signal with respect to the pump–probe delay.[58,50] In contrast, in F-PP, the 2Q coherence is modulated into the fluorescence signal by projecting it onto a final population via the last light–matter interaction, thus facilitating its detection.

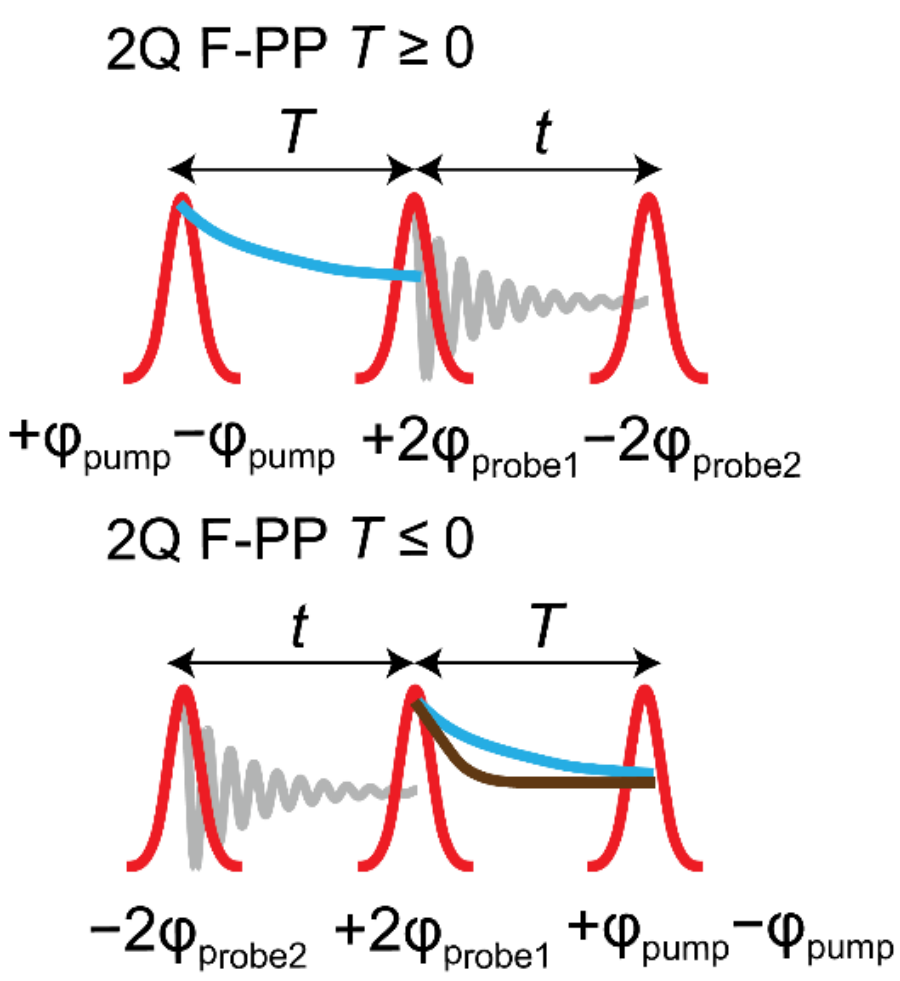

**Figure 3.** Two-quantum (2Q) F-PP spectroscopy. Pulse sequences with unique phase signatures described by the phase imparted from individual pump-pulse ($\varphi_{pump}$) and probe-pulse interactions ($\varphi_{probe1}$ and $\varphi_{probe2}$). A train of three pulses (red) is shown with mutual delays $T$ and $t$ and time ordering between pump and probe-pulse pairs for $T \geq 0$ (top) and $T \leq 0$ (bottom). We have 2Q coherences (gray) that evolve during $t$ for 2Q F-PP. For $T \leq 0$, population dynamics evolve between various doubly excited states (brown), while population evolves across singly excited states (blue) during both $T \leq 0$ and $T \geq 0$ delays.

The 2Q F-PP signal probes 2Q coherences along $t$ and thus requires each probe pulse to interact twice with the system in addition to the two interactions of the pump (Figure 3). This effectively results in a sixth-order signal with population dynamics evolving over $T$. The sixth-order nonlinear response is the leading-order contribution to the 2Q F-PP signal. In general, for an $M$Q signal, where $M = 1, 2, 3, \ldots$, the nonlinear response of order $2M + 2$ provides the leading term. In addition, there are higher-order terms that scale with $I_{pump}^{M}$, where $I_{pump}$ is the intensity of the pump pulse, that all add up to the total response. On the other hand, there are also $M$Q signals of at least an order of $2M + 2$ that scale with the intensity $I_{probe}^{M}$ of the identical probe pulses. These signals can be easily separated by an adequate phase-cycling scheme.[43,44] Since we are discussing here the sixth-order 2Q signal that scales with $I_{\text{pump}}^2$, it is necessary in the experiment to ensure that any eighth-order signal that scales with $I_{\text{pump}}^3$, or even higher-order 2Q signals that scale with $I_{\text{pump}}^{M>3}$ do not contribute substantially and ideally do not rise above the noise floor. In the present work, the pulse intensities were chosen such that the 2Q signal scales quadratically with pump intensity for our dynamic range, indicating that higher-order contributions remain negligible. Under these conditions, the signal is dominated by the desired sixth-order term. Alternatively, some variant of intensity cycling[59] can be implemented for F-PP that would enable the systematic separation of sixth-order 2Q signals. However, that is beyond the scope of the present work and will be described in a future study. In the reminder of this section, we assume that we only have to consider the sixth-order contribution wherever we discuss the 2Q signal.

Phase cycling separates the "$+\varphi_{\text{pump}} - \varphi_{\text{pump}} + 2\varphi_{\text{probe1}} - 2\varphi_{\text{probe2}}$" sixth-order signal from nonlinear signals with different phase signatures. However, to measure the desired transient response following pump excitation, chopping of the pump pulse is required to remove the overlapping "$+2\varphi_{\text{probe1}} - 2\varphi_{\text{probe2}}$" probe-only fourth-order signal component. Similar to 1Q F-PP, chopping allows us to collect data by turning the pump on and off in alternating measurements. By subtracting the signals obtained with the pump off from those with the pump on, the probe-only background can be effectively removed, isolating the pure sixth-order nonlinear response of interest. Phase cycling inherently takes care of the pump-only background signal during the extraction of the 2Q F-PP signal.

Three-pulse 2Q F-PP spectroscopy can be efficiently performed using 24-step nested [1×4×6] or 21-step cogwheel [COG21(0, 5, 6)] phase cycling[44] together with pump-pulse chopping. This pump chopping can be performed via a pulse shaper. In general, the F-PP technique can acquire arbitrarily high multi-quantum F-PP signals provided that the measurement is conducted with adequate phase-cycling schemes.[43,44]

To understand the dynamics encoded in the 2Q F-PP signal, we consider the same collection of subsystems as described earlier (Figure 1b and its generalization to $N$ subsystems), but now adapted to the 2Q case (Figure 4). We depict exemplary Feynman diagrams for $T \geq 0$ in Figure 4a and for $T \leq 0$ in Figure 4b. These correspond to the different types of pathways for the two-subsystem case of Figure 1b, but now associated with the sixth-order 2Q F-PP signals. The 2Q F-PP pathways for $T \geq 0$ include ground-state pathways ($Q_6$, $Q_7$, and $Q_8$), singly excited pathways ($Q_9$ and $Q_{10}$), and energy-transfer pathways ($Q_{11}$ and $Q_{12}$). The ground-state pathways, singly excited pathways, and energy-transfer pathways are classified following the same description as for 1Q F-PP. The dynamics for $T \leq 0$ in 2Q F-PP can be represented using ground-state pathways ($Q_6$', $Q_7$', and $Q_8$'), singly excited pathways ($Q_9$' and $Q_{10}$'), energy-transfer pathways ($Q_{11}$' and $Q_{12}$'), and doubly excited pathways ($Q_{13}$' and $Q_{14}$'). The doubly excited pathways are diagrams associated with the relaxation of a doubly excited state to the singly excited state during $T$. If we considered triply or higher excited states in the manifold of excited states (Figure 1b), we would need to account for additional pathways for both positive and negative $T$. These diagrams represent 2Q coherences that occur between the manifold of singly excited states and the manifold of triply excited states of the combined system during $t$. However, these pathways would cancel each other off, as we assume rapid relaxation from the higher excited manifolds to the singly excited manifold.

All the ground-state pathways, singly excited pathways, energy-transfer pathways, and doubly excited pathways can be further grouped as self-population (yellow shaded) or cross-population A (gray), B (pink), C (blue), and D (green) pathways, as depicted in Figures 4a and 4b. This distinction is based on the states between which coherence evolves and on the population dynamics evolving with $T$.

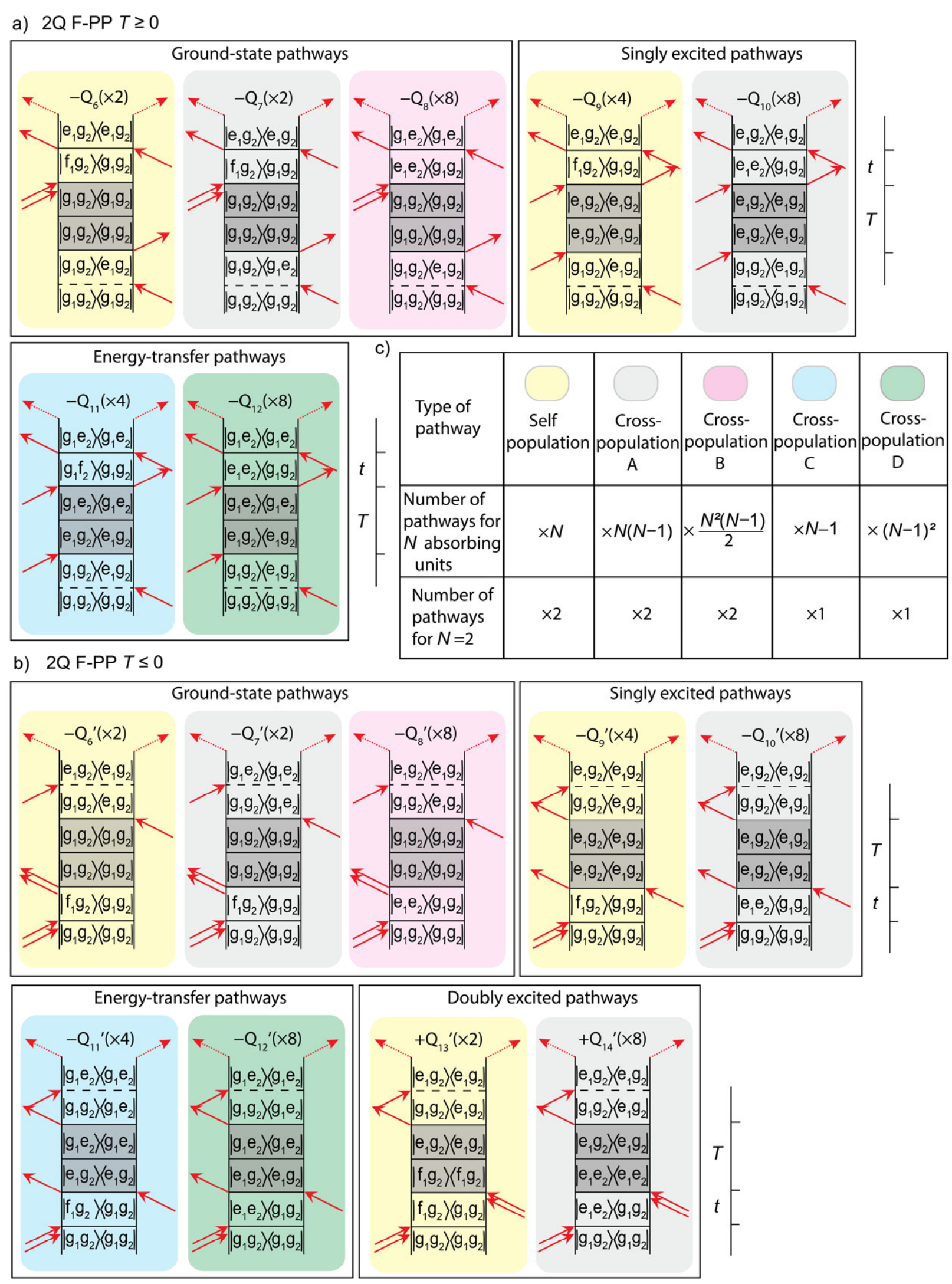

**Figure 4.** Double-sided Feynman diagrams for 2Q F-PP in the case of a dimer and generalization to larger systems. (a, b) Double-sided Feynman diagrams describing the 2Q F-PP signal with two subsystems and assuming fast relaxation from the doubly excited states to the singly excited states. The 2Q F-PP Feynman diagrams are shown for (a) $T \geq 0$ and (b) $T \leq 0$. Ground-state ($Q_6$, $Q_7$, $Q_8$, $Q_6$', $Q_7$', and $Q_8$') and singly excited ($Q_9$, $Q_{10}$, $Q_9$', and $Q_{10}$') pathways represent diagrams for which the population of the system evolves in the ground and singly excited state, respectively. Energy-transfer pathways ($Q_{11}$, $Q_{12}$, $Q_{11}$', and $Q_{12}$') show energy transfer between singly excited states. All pathways except the doubly excited pathways ($Q_{13}$' and $Q_{14}$') and energy-transfer pathways cancel off for the difference spectrum Δ2Q F-PP ($\omega_t$, $T$). The weights indicated in parentheses above each pathway result from different possible time-ordering sequences of light–matter interactions and the possibilities to generate a final excited-state population that leads to fluorescence. (c) The pathways can be categorized as self-population (yellow-shaded) and cross-population pathways, based on the number of subsystems required for a particular pathway. The cross-population pathways A (gray shaded), B (pink shaded), C (blue shaded), and D (green shaded) labels are used for categorizing pathways for a system containing $N$ subsystems. The number of pathways is quoted in each case for arbitrary $N$ (middle row) and $N = 2$ (last row).

In the 2Q F-PP signal for $T \geq 0$ (Figure 4a), the self-population pathways involve the initial pump excitation to create a coherence between a singly excited state and the ground state. In 2DES, the technique's ability to resolve coherences along the $\tau$ axis, representing the coherence time between the first two pump pulses, enables the detection of such coherences between electronic states. This coherence cannot be resolved in a 2Q F-PP experiment, unlike in 2DES, because of the lack of the additional coherence delay axis. Following the complete two electric-field interactions of the pump pulse on one of the two subsystems, the system populates the ground state ($Q_6$) or a singly excited $|E\rangle$ state ($Q_9$) during $T$. Finally, the interaction of the probe pulses generates a 2Q coherence that evolves during $t$, between the $|f_i\rangle$ state and the ground state of the same subsystem the pump interacts. It is important to note that in Figures 4a and 4b, we present only one exemplary pathway for each pathway type. However, multiple $Q_n$ pathways can be drawn for a given value of $n$ ($n$ = 1, ..., 14), based on our description. In the ground-state cross-population A ($Q_7$) and energy-transfer cross-population C ($Q_{11}$) pathways, the initial pump interactions take place on one of the subsystems. The probe pulses, however, create an excitation on the other subsystem (labelled with a different value of $i$) during the $t$ delay by creating a 2Q coherence between the ground state and an $|f_i\rangle$ state of this second subsystem. Meanwhile, the ground-state cross-population B ($Q_8$), singly excited cross-population A ($Q_{10}$), and energy-transfer cross-population D ($Q_{12}$) pathways involve a 2Q coherence between the ground state and $|e_1e_2\rangle$ as a result of the electric field interactions of the probe pulses occurring simultaneously on both subsystems.

In the 2Q F-PP signal for $T \leq 0$ (Figure 4b), the first probe pulse interacts with the system creating a 2Q coherence that evolves as a function of $t$. The second probe pulse creates a population in the ground, singly excited, or doubly excited state, with the final two pump interactions leading to a fluorescence emission that is detected. For each pathway type in 2Q F-PP for $T \geq 0$, there is a corresponding pathway for $T \leq 0$ that represents the same population dynamics as a function of $T$ and encompasses a coherence between the same electronic states that oscillates as a function of $t$. For each ground-state and singly excited $Q_n$ pathway for $T \geq 0$, the corresponding pathway for $T \leq 0$ is labeled as $Q_n$'. The only pathways that are exclusive to the dynamics of 2Q F-PP for $T \leq 0$ are the doubly excited self-population pathways ($Q_{13}$') and the doubly excited cross-population A pathways ($Q_{14}$'). In the cross-population A pathways, the 2Q coherence evolves between the ground state and the $|e_1e_2\rangle$ state, with a simultaneous excitation of both subsystems. Meanwhile, the doubly excited self-population pathways contain a coherence between the ground and an $|f_i\rangle$ state.

The energy-transfer cross-population C and D pathways along $T \geq 0$ and $T \leq 0$ represent the energy transfer between the singly excited states. The energy-transfer dynamics contribution in the 2Q F-PP map from the cross-population C pathways are detected at distinct frequencies along $\omega_t$ for both positive and negative sections of the $T$ delay. This is different from all the ground- and singly excited-state pathways where there is one $Q_n$ pathway for each $Q_n$' pathway. This difference occurs because the electronic states between which coherence evolves during $t$ is different for $T \geq 0$ and for $T \leq 0$ in these energy-transfer pathways. In the case of cross-population C pathways, we have a coherence between ground and an $|f_i\rangle$ state of one subsystem ($Q_{11}$) for $T \geq 0$, while the coherence evolves across the ground and an $|f_i\rangle$ state of the other subsystem ($Q_{11}$') for $T \leq 0$.

The cross-population pathways $Q_7$, $Q_8$, $Q_{10}$, $Q_{11}$, and $Q_{12}$ in 2Q F-PP for $T \geq 0$ (Figure 4a) and $Q_7$', $Q_8$', $Q_{10}$', $Q_{11}$', $Q_{12}$', and $Q_{14}$' in 2Q F-PP for $T \leq 0$ (Figure 4b) describe incoherent mixing. Incoherent mixing in fourth-order nonlinear experiments can be expressed as a product of two second-order linear signals originating from two different subsystems modulated with the same phase signature as the regular fourth-order self-population signal during the temporal window of fluorescence detection. Analogously, incoherent mixing in sixth-order experiments is conceptualized as the mixing of a fourth-order nonlinear signal and a second-order linear signal originating from two different subsystems during fluorescence detection. The concept of incoherent mixing in sixth-order 2Q F-PP can be extended from a dimer to a system with $N$ subsystems, and we can quantify the number of ground-state, singly excited, energy-transfer and doubly excited pathways.

We consider a three-manifold electronic state structure for a system consisting of $N$ subsystems, similar to the one described for the dimer in Figure 1b. Each subsystem individually contains a ground state $|g_i\rangle$, a singly excited state $|e_i\rangle$, and a doubly excited state $|f_i\rangle$, where the subscript, $i$ = 1, ..., $N$, labels the subsystem. In the combined system, there are $N$ singly excited states in the $|E\rangle$ manifold, $N$ doubly excited states corresponding to the individual double excitation to $|f_i\rangle$ of only one of the subsystems, and $\frac{N(N-1)}{2}$ additional states in the manifold of $|F\rangle$ arising from the simultaneous excitation of the $|e_i\rangle$ states of two different subsystems.

We can tabulate the number of self-population and cross-population A, B, C, and D pathways by tracking the number of subsystems excited by the pulses and the states involved in this process. Since self-population pathways (e.g., $Q_6$, $Q_9$, and $Q_{13}$',) require the excitation of the $|e_i\rangle$ and $|f_i\rangle$ states of the same subsystem during the sequence of electric-field interactions, we will have $N$ self-population pathways. In the cross-population A ground-state pathways (e.g., $Q_7$), the system resides in $|G\rangle$ during $T$ after exciting one of the subsystems and the sequence of field interactions during $t$ results in excitations on a different subsystem. Specifically, if one of the $N$ subsystems is excited before the system returns to the ground state, there are $(N-1)$ other subsystems that can be excited during $t$. As a result, the total number of such cross-population pathways is given by the combination $N(N-1)$. In cross-population B ground-state pathways (e.g., $Q_8$), each of the $N$ subsystems can be independently excited and subsequently de-excited during the evolution period $T$. After de-excitation, the combined system returns to and populates the ground state $|G\rangle$. Following this, a coherence can be established between $|G\rangle$ and any of the $\frac{N(N-1)}{2}$ states in $|F\rangle$ arising from the simultaneous excitation of the $|e_i\rangle$ states of two different subsystems. Since there are $N$ choices for the initially excited-and-deexcited subsystem and $\frac{N(N-1)}{2}$ possible pairs of distinct subsystems that can be simultaneously excited to form $|F\rangle$, the total number of such coherence pathways is $\frac{N^2(N-1)}{2}$.

Cross-population A singly excited-state pathways (e.g., $Q_{10}$) correspond to cases where the system is in a state in $|E\rangle$ formed from exciting one of the $N$ subsystems during $T$, and during $t$, a coherence is formed between $|G\rangle$ and a state in $|F\rangle$ involving double excitation, with one excitation on the same subsystem excited during $T$, and the other on a different subsystem. For each of the $N$ states in $|E\rangle$ that can be populated during $T$, $(N-1)$ coherences using these doubly excited states in $|F\rangle$ can exist, yielding a total of $N(N-1)$ pathways. Cross-population A doubly excited-state pathways (e.g., $Q_{14}$) correspond to cases where the system is populating the state in $|F\rangle$ formed from the double excitation of $|e_i\rangle$ states of any two of the $N$ subsystems, followed by relaxation to any one of the states in $|E\rangle$ during $T$. During $t$, a coherence exists between this same state in $|E\rangle$ and $|G\rangle$. Since such a diagram can exist with any of the $N$ states in $|E\rangle$ during $t$, and by considering $(N-1)$ states in $|F\rangle$ formed from double excitation of $|e_i\rangle$ states during $T$, we have a total combination of $N(N-1)$ cross-population A doubly excited-state pathways. In cross-population C energy-transfer pathways (e.g., $Q_{11}$), a total of $(N-1)$ pathways can be represented by describing the relaxation of $(N-1)$ higher-energy states in the manifold of $|E\rangle$ populated during $T$ to the lowest-energy $|E\rangle$ state. In this case, the 2Q coherence along $t$ evolves between $|G\rangle$ and a state in $|F\rangle$ formed due to double excitation of the same lowest-energy subsystem. Cross-population D energy-transfer pathways (e.g., $Q_{12}$) describe the relaxation of $(N-1)$ higher-energy states in $|E\rangle$ populated during $T$ to the lowest-energy state in the $|E\rangle$ manifold. During $t$, excitations create a coherence that evolves between $|G\rangle$ and the state in $|F\rangle$ formed due to double excitation of the lowest-energy subsystem along with any of the $(N-1)$ higher-energy subsystems. This yields a total of $(N-1)^2$ pathways.

For every pathway labelled with $Q_n$ or $Q_n$', we will also have different possible time-ordering sequences of light–matter interactions. In addition, we account for the number of possibilities in which a final excited-state population that gives rise to fluorescence can be generated. The product of the number of time-ordering sequences and the number of final excited-state configurations corresponds to the number of times each individual pathway contributes to the response. These numbers are given as "weights" in parentheses above each pathway in Figures 4a and 4b. These weights were obtained by automated counting, using the ultrafast spectroscopy suite software developed by Krich et al.[60,61] Then, by combining these weights with the numbers shown in Figure 4c, we obtain a combined total of $4N^3 - 2N^2$ ground-state pathways, $8N^2 - 4N$ singly excited pathways, $8N^2 - 12N + 4$ energy-transfer pathways, and $8N^2 - 6N$ doubly excited pathways, where self- and cross-population contributions were added for each pathway type. Now consider an asymptotic analysis in the limit of large $N$, where for each pathway type the total number of pathways is approximated by the highest-order term in the corresponding sum of Figure 4c. The ratio of the number of ground-state pathways to excited-state pathways gives insight into how incoherently mixed signals affect the 2Q F-PP spectrum. In particular, the ground-state pathways scale by a factor of $N$ relative to other pathway contributions that contain population dynamics of excited states. Thus, incoherent mixing should also be considered as a severe issue in the sixth-order spectroscopy of large systems, as the signal is dominated by static background ground-state pathways for large $N$.

The asymmetry in the distribution of doubly excited pathways for $T \leq 0$ compared to $T \geq 0$ in addition to the symmetry of other pathways relative to $T = 0$ allows the selective extraction of dynamics associated with the doubly excited states by performing a subtraction like that described for 1Q F-PP in Section 2.1. The difference signal, Δ2Q F-PP ($\omega_t$, $T$), obtained analogously and defined for $T \geq 0$, is given by:

$$\Delta 2\text{Q F-PP}\ (\omega_t, T) = 2\text{Q F-PP}\ (\omega_t, T) - 2\text{Q F-PP}\ (\omega_t, -T). \quad (2)$$

The subtraction removes all ground-state and singly excited pathways and leaves only the doubly excited pathways ($Q_{13}$' and $Q_{14}$') and the energy-transfer pathways ($Q_{11}$, $Q_{12}$, $Q_{11}$', and $Q_{12}$') in the Δ2Q F-PP ($\omega_t$, $T$) difference spectrum. As for the case of 1Q F-PP, this subtraction procedure only works in case of a Markovian bath; otherwise the symmetry of undesired contributions relative to $T = 0$ will be altered. The removal of ground-state pathways effectively eliminates incoherent mixing in the difference spectrum, leaving only the dynamics of the excited states. This is especially advantageous for systems with a large number of chromophores, where dynamical processes are otherwise strongly obscured by ground-state pathways in the ordinary F-PP spectra.

Note that the large background arises because, in systems with $N \gg 2$, the ground-state pathways scale more strongly with $N$ than any other pathway type due to the increasing number of permutations arising from pump and probe interactions taking place on different sites.

Extending the analysis of 2Q F-PP to the case including electronic coupling between subsystems, the doubly excited product states mix to form coupled multiexcitonic states, thus facilitating annihilation from the doubly excited manifold. The manifold contains both biexciton and mixed states in $|\mathrm{F}\rangle$ formed due to double excitation of the same subsystem, that relax to the excitonic states formed by coupled $|\mathrm{E}\rangle$ states. The difference signal Δ2Q F-PP ($\omega_t$, $T$) in the case of an excitonic system hence contains information about dynamics represented by doubly excited and energy-transfer pathways between excitonic states.

The information on exciton dynamics encoded in sixth-order 2Q F-PP can be inferred from the theoretical description of sixth-order nonlinear signals in literature, described in the framework of 2DES.[6] The relation between 2Q F-PP and the sixth-order fluorescence-detected exciton–exciton-interaction two-dimensional (EEI2D) signal[6] can be explained via the projection-slice theorem.[62] This theorem describes how the pump–probe spectrum can be retrieved from the projection of the real part of the absorptive 2D spectrum onto the detection axis for all $T$ delays between the pump and the probe pulse pairs. In sixth-order EEI2D spectroscopy, the 1Q–2Q and 2Q–1Q signal components are separated from each other using phase cycling. While the 1Q–2Q signal contains 1Q coherence evolving along the excitation and 2Q coherence along the detection interpulse delay of the 2D experiment, it is vice versa for the 2Q–1Q signal. These 2Q–1Q and 1Q–2Q signals both probe population dynamics over $T$. The double Fourier transformation of 1Q and 2Q coherences along the interpulse delays helps to resolve time-domain data within the frequency domain by creating separate $\omega_{1\mathrm{Q}}$ and $\omega_{2\mathrm{Q}}$ frequency axes for the 2D spectrum for each value of $T$, resulting in $R_{1Q-2Q}(\omega_{1Q},\ \omega_{2Q},\ T)$ and $R_{2Q-1Q}(\omega_{2Q},\ \omega_{1Q},\ T)$ 3D datasets. The projection of the 1Q–2Q 2D spectrum from $R_{1Q-2Q}(\omega_{1Q}, \omega_{2Q},\ T)$ data onto the 2Q axis for each *T*, yields the $R_{1Q-2Q}(\omega_{2Q},\ T)$ spectra, given by:

$$R_{1Q-2Q}(\omega_{2Q},\ T) = \int \left[R_{1Q-2Q}(\omega_{1Q},\ \omega_{2Q},\ T)\right] \mathrm{d}\omega_{1Q}. \quad (3)$$

The $R_{1Q-2Q}(\omega_{2Q},\ T)$ spectra is mathematically equivalent to the 2Q F-PP spectrum for $T \geq 0$, provided that the $\omega_t$ 2Q F-PP detection axis is the same as the $\omega_{2Q}$ 2D axis. The 2Q F-PP spectrum for $T \leq 0$, on the other hand, under the same condition of $\omega_{2Q} = \omega_t$, is equivalent to $R_{2Q-1Q}(\omega_{2Q},\ T)$, obtained by the projection of the 2Q–1Q 2D spectrum from $R_{2Q-1Q}(\omega_{2Q},\ \omega_{1Q},\ T)$ onto the 2Q axis,

$$R_{2Q-1Q}(\omega_{2Q},\ T) = \int \left[R_{2Q-1Q}(\omega_{2Q},\ \omega_{1Q},\ T)\right] \mathrm{d}\omega_{1Q}. \quad (4)$$

These two maps are then subtracted from each other to get the sixth-order 2D difference signal ΔF-EEI2D ($\omega_{2Q}$, *T*),

$$\Delta\text{F-EEI2D}\ (\omega_{2Q}, T) = R_{1Q-2Q}(\omega_{2Q},\ T) - R_{2Q-1Q}(\omega_{2Q},\ T). \quad (5)$$

This process is analogous to the evaluation of Δ2Q F-PP ($\omega_t$, *T*) in Eq. (2) and becomes equivalent to ΔF-EEI2D ($\omega_{2Q}$, *T*) under the condition of $\omega_{2Q} = \omega_t$. From an experimental point of view, using sixth-order 2Q F-PP is more practical than sixth-order fluorescence-detected EEI2D spectroscopy in terms of the required measurement time. While sixth-order EEI2D spectroscopy requires 125-step nested phase cycling for extracting the required nonlinear signal contributions[6], the 2Q F-PP experiments employs a significantly reduced 24-step nested phase cycling scheme.

Moreover, despite the requirement of chopping, the absence of a coherence time further leads to a substantial reduction in measurement time.

The Δ2Q F-PP ($\omega_t$, $T$) spectrum contains energy transfer and annihilation in a coupled system, which can in principle be analyzed with global analysis.[63,64] In certain systems however, there could be overlapping and convoluted spectral features in the spectrum that complicate its interpretation. To describe a unified trend in the dynamics described by Δ2Q F-PP ($\omega_t$, $T$) for all multichromophore systems, we consider a new quantity Δ2Q F-PP ($T$),

$$\Delta 2\text{Q F-PP }(T) = \int \Delta 2\text{Q F-PP }(\omega_t, T)\ \mathrm{d}\omega_t, \tag{6}$$

obtained by integrating Δ2Q F-PP ($\omega_t$, $T$) over $\omega_t$. It rises according to the rate of energy transfer and annihilation in the coupled system. Similarly, we introduce the frequency-integrated 1Q analog,

$$\Delta 1\text{Q F-PP }(T) = \int \Delta 1\text{Q F-PP }(\omega_t, T)\ \mathrm{d}\omega_t, \tag{7}$$

Let us now discuss Δ2Q F-PP ($T$) in a dimer for two different cases. In the first case, we assume identical transition dipole moments $\mu_{A0}$ and $\mu_{B0}$, corresponding to transitions from the ground state $|0\rangle$ to the higher and lower single-exciton states $|\mathrm{A}\rangle$ and $|\mathrm{B}\rangle$, respectively. For this system, the Δ2Q F-PP ($T$) signal rise would correspond exclusively to the doubly excited annihilation dynamics without any energy-transfer contribution. This is because the energy-transfer pathways would make the same contribution in case of identical transition dipole moments but carry opposite signs in Δ2Q F-PP ($\omega_t$, $T$) and thus cancel off in Δ2Q F-PP ($T$). In the second case, with $\mu_{A0} \neq \mu_{B0}$, Δ2Q F-PP ($T$) contains, in addition to annihilation denoted by the rate $k_A$, single-exciton transfer denoted by the rate $k_T$. The reason behind that is incomplete cancellation of the energy-transfer pathways upon signal integration, as the energy-transfer pathways have different weights due to the different transition dipole moments. This circumstance was previously[6] illustrated by the population-time propagator of the ΔF-EEI2D ($\omega_t$, $T$) signal under impulsive excitation. The spectrally integrated propagator of the Δ2Q F-PP signal is given by

$$\begin{aligned}\Delta\text{2Q F-PP}\,(T) \propto \Big\{&-4(\mu_{A0}^4\mu_{B0}^2 + \mu_{A0}^2\mu_{B0}^4) \\ &+ 4(\mu_{A0}^4\mu_{B0}^2 - \mu_{B0}^4\mu_{A0}^2)\left(1 + \frac{k_A}{-k_A + k_T}\right)\exp(-k_T T) \\ &+ 4\mu_{B0}^2\mu_{A0}^2\left[\mu_{B0}^2\left(1 + \frac{k_T}{-k_A + k_T}\right) - \mu_{A0}^2\frac{k_A}{-k_A + k_T}\right]\exp(-k_A T)\Big\}.\end{aligned} \tag{8}$$

The second term of Eq. (8) describes the contribution of single-exciton transfer that decays exponentially according to rate $k_T$. Its amplitude scales with the difference $\mu_{A0}^4\mu_{B0}^2 - \mu_{B0}^4\mu_{A0}^2$, which results in a finite contribution in the case of $\mu_{A0} \neq \mu_{B0}$. Note that, because of the close relation of ΔF-EEI2D ($\omega_{2Q}$, $T$) and Δ2Q F-PP ($\omega_t$, $T$) described previously, their propagators are the same.[6] According to Eq. (8), the ratio between single-exciton transfer and annihilation that enters Δ2Q F-PP ($T$) is determined by the ratio of $\mu_{A0}$ and $\mu_{B0}$. In the case of a homodimer, the ratio of the transition dipole moments of the resulting exciton states is thus crucial as well. In addition, the overlap of the laser spectrum with the exciton bands affects how much single-exciton transfer is contained in Δ2Q F-PP ($T$), as further discussed in Section 3.2. Furthermore, the analysis can be correspondingly extended to the case of polymers. The squaraine heteropolymers discussed in Section 3.2 also feature two prominent exciton bands, thus, in that case, $\mu_{A0}$ and $\mu_{B0}$ can be considered as effective transition dipole moments. Noteworthily, the removal of incoherent mixing contributions upon the construction of Δ2Q F-PP ($\omega_t$, $T$) works for all types of multichromophoric systems and is not limited to heterogeneous systems. In general, one can extract energy transport and annihilation times in homogeneous systems provided there is a possibility for excitons to propagate in the system. While homogeneous systems often feature broad spectral features that cannot be attributed to individual participating chromophores, higher-order methods such as fifth-order C-PP[59] and sixth-order 2Q F-PP, as presented in this work, yield characteristic timescales for energy transport as these methods are capable of directly probing the event of exciton interaction after diffusion.

A benefit that 2Q F-PP provides over coherently detected higher-order PP spectroscopy[59,65] for the extraction of doubly excited annihilation dynamics is that Δ2Q F-PP ($\omega_t$, $T$) removes contaminations from so-called parasitic multi-quantum signals originating from incorrect time-ordered interactions at pulse overlap.[47,48] These signals usually perturb the analysis of higher-order signals. The “desired” 2Q F-PP nonlinear response is based on well time-ordered interactions for all delay steps. However, if finite pulses are considered, pathways that result from wrong time ordering (parasitic pathways) contaminate the desired nonlinear response on a time scale that is comparable to the employed pulse duration.[47,48] These parasitic pathways arise from the overlap of two or more pulses in the 2Q F-PP three-pulse excitation sequence. The time ordering of the pulse interactions must be considered for a correct determination of the individual pathway contributions in the sixth-order 2Q F-PP signal.

We will have many combinations of interactions that carry the same total phase signature but generate different dynamics (Figures 5a and 5b). The overlap between the pump pulse (green Gaussian) and the first probe pulse (red Gaussian) [shown in the right halves of Figures 5a ($T \geq 0$) and 5b ($T \leq 0$)] as well as the overlap between the two pulses of the probe-pulse pair [red and blue Gaussians, depicted in the left halves of Figures 5a ($T \geq 0$) and 5b ($T \leq 0$)] leads to these parasitic pathways, in addition to the correctly time-ordered pathways (Figures 4a and 4b). When all the pulses overlap simultaneously, we need to take care of all these pathways in the analysis.

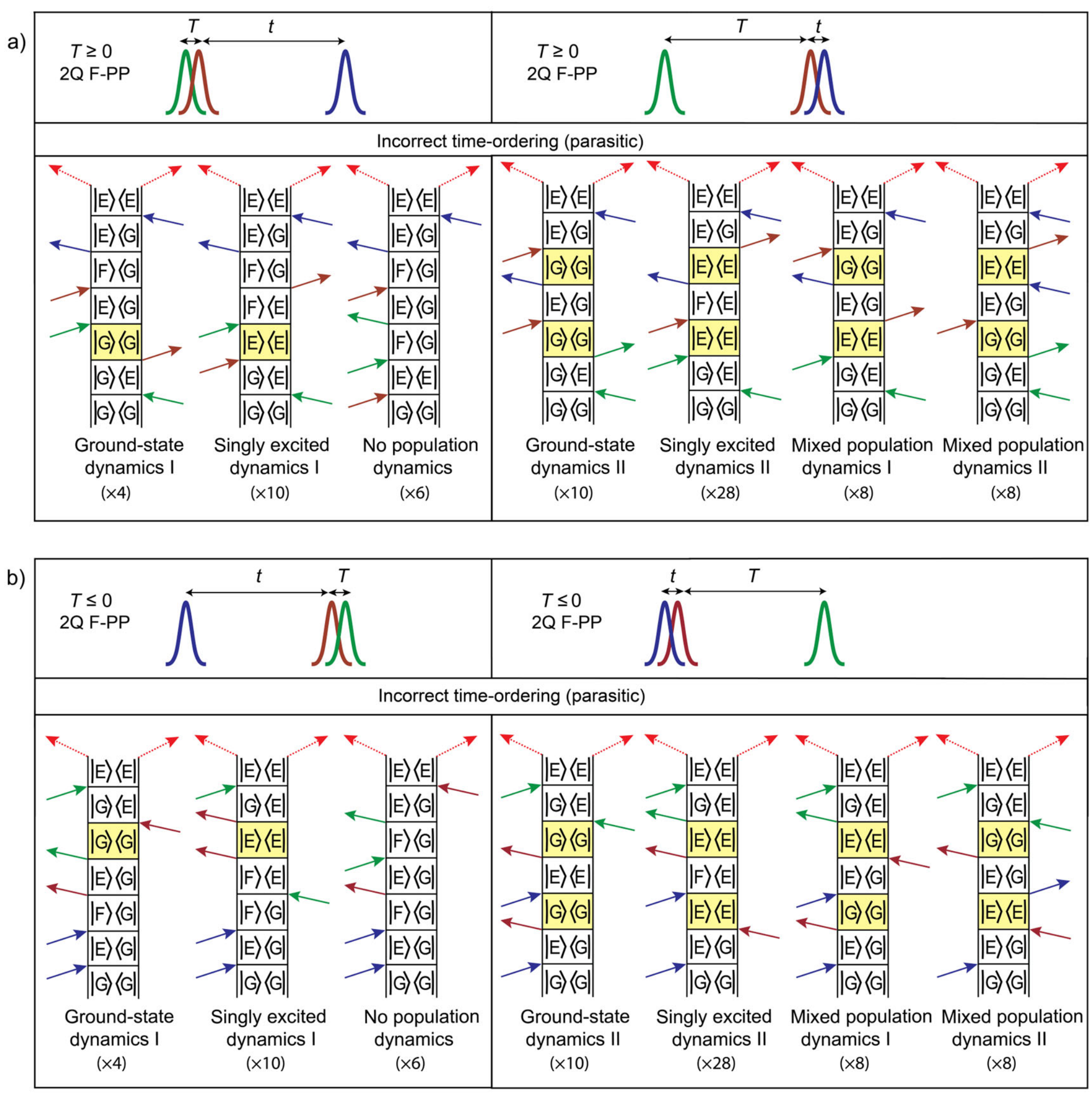

**Figure 5.** Double-sided Feynman diagrams for 2Q F-PP during pulse overlap. We consider a system with three manifolds of electronic states, a ground state |G⟩, a first excited state |E⟩, and a second excited state |F⟩ and assume fast annihilation from |F⟩ to |E⟩. The 2Q F-PP diagrams are shown (a) for $T \geq 0$ and (b) for $T \leq 0$. For both cases, the Feynman diagrams for correct time-ordered interactions are already shown in Figure 4. The incorrect time ordering between pump–probe and probe–probe interactions results in different types of pathways. The pump pulse and associated interactions are represented by green arrows in the diagrams. The pulse interactions with the first and the second pulse in the probe pair are represented by red and blue arrows, respectively. The weight of each pathway type, i.e., the number of times it contributes to the response, is provided at the bottom. These numbers were obtained using the ultrafast spectroscopy suite software developed by Krich et al.[60,61] The diagrams are classified by the population

dynamics they involve, which are highlighted with a yellow shaded area. Pathways labeled with "ground-state dynamics" are associated with the evolution of the ground-state population. Pathways labeled with "singly excited dynamics I, II" and "mixed population dynamics" also contain population dynamics of singly excited states. When all pulses overlap in time, we get a combination of all these pathways simultaneously.

These parasitic contributions are a problem in studying the dynamics of multiply excited states using coherently detected higher-order PP and 2D spectroscopy. The overlapping signal contributions overwhelmingly carry the dynamics of singly excited states and contaminate the doubly excited-state dynamics expected to be probed with the higher-order signal.[6,59,66] Consequently, singly excited-state dynamics will mix with annihilation, especially at early $T$. Thus, the finite duration of ultrashort pulses limits the retrieval of annihilation kinetics in a coherently detected approach. This problem is particularly severe in cases where the multi-quantum coherences dephase on a time scale similar to the pulse duration. In such a case, the distinction between parasitic and "real" multi-quantum coherences is extremely difficult. To minimize this problem, the pulse duration should therefore be as short as possible, which may, however, pose an experimental limitation.[47,48]

The acquisition of the 2Q F-PP signal for both positive and negative $T$ delays and the subsequent post-processing to create the difference signal Δ2Q F-PP ($\omega_t$, $T$) via Eq. (2) fully eliminates the contamination by parasitic pathways. This is because parasitic pathways are distributed symmetrically along the $T \geq 0$ and $T \leq 0$ sections of the 2Q F-PP map. All parasitic pathway types can be classified based on the population states that are involved during light–matter interaction. They can be categorized into ground-state I and II pathways, for which the population evolves in the ground state, and as singly excited-state I and II pathways, for which the population evolves in the singly excited state. The distinction between I and II pathways refers to how many times the system populates the respective state – either once (I) or twice (II), during the sequence of field interactions. In the case of ground-state pathways, this means the ground state is populated once (I) or twice (II). Similarly, for singly excited-state pathways, the singly excited state is populated once (I) or twice (II). We also have parasitic mixed population dynamics I and II pathways, for which the system can exist in both ground and singly excited states at different instances within the interval of multi-pulse excitation. For $T \geq 0$, the I and II distinction indicates which state was populated first during the sequence of pulse excitations: singly excited state for pathway I, and ground state for pathway II. For $T \leq 0$, the meaning is reversed: the ground state is populated first in pathway I, while the singly excited state is populated first in pathway II. The number of each of these parasitic pathways remains the same across the $T \geq 0$ and $T \leq 0$ delay sections. All parasitic pathways can therefore be removed by calculating the difference signal, allowing extraction of the doubly excited-state dynamics.

## 3. Results and discussion

### 3.1. Simulation

To illustrate the general properties of 1Q and 2Q F-PP spectra for both positive and negative $T$, we consider a "minimal" and exemplary homodimer model in the exciton basis with a ground state, two single-exciton states at energies of 1.73 eV and 1.88 eV, and a biexciton state at 3.61 eV. We conducted non-perturbative dynamics simulations in the Lindblad formalism using our MATLAB toolbox QDT.[67] In the simulation, we used three-pulse sequences of Gaussian pulses with 10 fs duration centered at 1.79 eV. The time of energy transfer between the single-exciton states was set to 100 fs, while annihilation was effectively modeled with a relaxation of the biexciton state to the upper single-exciton state, also with 30 fs relaxation time. The dephasing time of the inter-excitonic coherence between the two single-exciton states was intentionally chosen so that coherent oscillations are visible in the 1Q and 2Q F-PP spectra. The pump pulse was chopped and 5×5-fold phase cycling of the probe pulses was performed to resolve the 1Q and 2Q F-PP signals without aliasing. These two signals were then extracted by first subtracting the data without the pump pulse from those with the pump pulse and by applying suitable phase weights.[43] A Fourier transformation of the data along $t$ then yielded the 1Q and 2Q F-PP spectra. Further simulation parameters are listed in the SI.

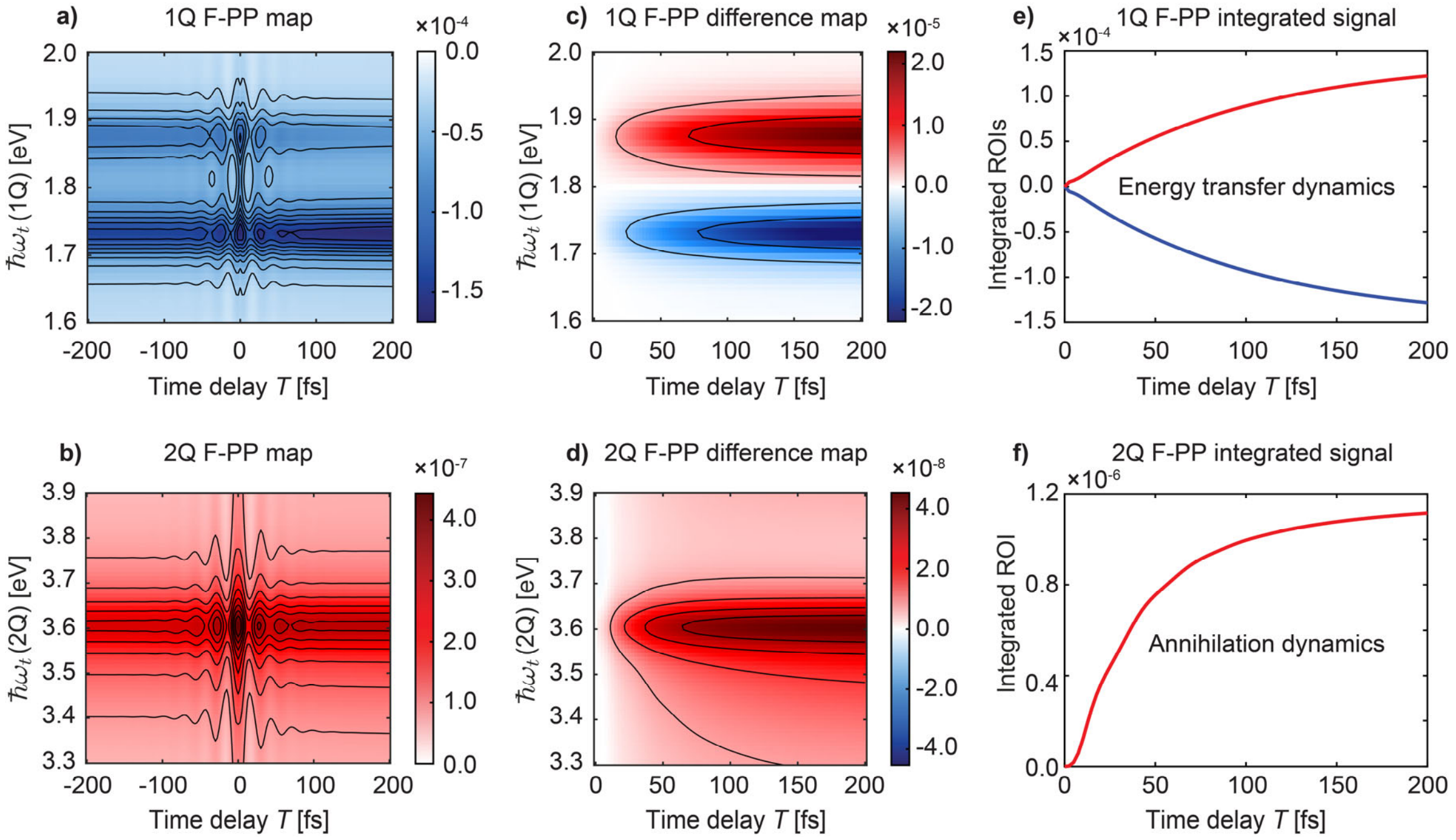

**Figure 6.** Simulation of one-quantum (1Q) and two-quantum (2Q) F-PP spectroscopy for an exemplary dimer system with a ground state, two single-exciton states, and a biexciton state. (a) 1Q F-PP map. (b) 2Q F-PP map. (c) Δ1Q F-PP ($\omega_t$, $T$) map according to Eq. (1). Opposite-signed features appear at the energies of the two single-exciton states. (d) Δ2Q F-PP ($\omega_t$, $T$) map according to Eq. (2). (e) 1Q F-PP signal obtained via the spectral integration of Δ1Q F-PP ($\omega_t$, $T$) spectrum over each single-exciton feature in the difference map. The lower peak is integrated from 1.71 eV to 1.76 eV to yield the blue curve, while the upper peak is integrated from 1.85 eV to 1.90 eV to yield the red curve. The integrated signal as a function of $T$ decays or rises, respectively, with the rate of energy transfer between the single-exciton states. (f) 2Q F-PP signal obtained via the spectral integration of Δ2Q F-PP ($\omega_t$, $T$) over the complete energy axis. The signal rises with the rate of the annihilation of the biexciton to the single-exciton states.

The negative and positive regions of the F-PP maps (1Q in Figure 6a and 2Q in Figure 6b) are separated by a “coherent spike” associated with spatial and temporal pulse overlap between all pulses including the pump and the probe pulse pair.[1] The two transient bands in the 1Q F-PP map energetically correspond to the two single-exciton states of the dimer that was simulated, while the single broad band across the 2Q F-PP detection axis represents the biexciton state. Subtracting the positive F-PP ($T \geq 0$) (Figure 6a) and negative F-PP ($T \leq 0$) (Figure 6b) delay sections of the transient maps from each other according to Eqs. (1) and (2) yields the Δ1Q F-PP ($\omega_t$, *T*) and Δ2Q F-PP ($\omega_t$, *T*) spectra (Figures 6c and 6d, respectively). In Δ1Q F-PP ($\omega_t$, *T*), two spectral features appear with opposite signs, originating from the $Q_5$ and $Q_5$' pathways discussed above (see also Figure 2). The $Q_5$ and $Q_5$' pathways both represent the dynamics of energy transfer from a higher-energy singly excited state to the lower-lying singly excited state along $T$. The $Q_5$' and Q5 pathways energetically correspond to the higher- and lower-energy features in Δ1Q F-PP ($\omega_t$, *T*), respectively (Figure 6c). The subtraction in Eq. (1), to obtain the Δ1Q F-PP ($\omega_t$, *T*), introduces a negative sign in the contribution of the $Q_5$' pathway. Integrating spectrally over the energetic region in $\omega_t$ corresponding to each of these features and plotting the result as a function of *T* results in a Δ1Q F-PP (*T*) signal that either decays (feature corresponding to lower energy) or rises (feature corresponding to higher energy) with the rate of energy transfer (Figure 6e). The Δ2Q F-PP (*T*) signal meanwhile almost entirely rises with the rate of annihilation (Figure 6f). Note that residual components associated with single-exciton contributions, due to unequal transition dipole moments of the single-exciton states [as discussed in Section 2.2, see also Eq. (8)], manifest themselves as a slight asymmetry in the lower-energy region of the difference signal in Fig. 6d. The occurrence of this asymmetry also means that the Δ2Q F-PP (*T*) transient contains a minor proportion of the 100-fs energy-transfer kinetics. However, by comparing Figs. 6e and 6f, it becomes clear that the fraction of energy-transfer dynamics is negligible in Δ2Q F-PP (*T*). In general, the proportion of energy-transfer dynamics in Δ2Q F-PP (*T*) can be minimized by adjusting the overlap of the laser spectrum with the linear absorption of the system so that the amplitudes of the two single-exciton features are the same in the excitation spectrum. It is worth noting that, in contrast to the F-PP maps, all the obtained difference maps show only subtle modulations along $T$ if noticeable at all, thus coherent oscillations due to inter-excitonic coherence between the two single-exciton states are almost perfectly eliminated in both Δ1Q F-PP ($\omega_t$, $T$) and Δ2Q F-PP ($\omega_t$, $T$), indicating that the subtraction strongly emphasizes the pure population dynamics in both cases. Moreover, due to its symmetry at $T = 0$, the coherent spike is eliminated by the subtraction as well,[51] leading to signals that rise or decay from zero in the difference spectra of both signals.

### 3.2. Experiment

We now illustrate 1Q and 2Q F-PP spectroscopy exemplarily in an experiment on the squaraine heterodimer [SQA–SQB] and the copolymer $[SQA–SQB]_5$.[66,68] The multiexciton dynamics of these squaraine systems have been a topic of investigation in previous coherently detected and fluorescence-detected 2D experiments.[6,66,69] The heterodimer exhibits J-type coupling with excitons partially localized on the SQA and SQB units. The energy-transfer and annihilation dynamics of this system were investigated using fourth-order F-PP, fourth-order 2DES, and sixth-order F-EEI2D experiments.[1,6,51,69] Hence, we can compare our F-PP method to those prior results. The copolymer was chosen as a second exemplary system that is subject to incoherent mixing due to the larger number of chromophores. Thus, we can illustrate the removal of incoherent mixing upon calculating the Δ1Q F-PP ($\omega_t$, $T$) and Δ2Q F-PP ($\omega_t$, $T$) spectra.

The 1Q F-PP map for the dimer (Figure 7a) displays two features appearing at 1Q energies that correspond to those of the two single-exciton states in the absorption spectrum (see SI). Along the 1Q axis, a dynamic Stokes shift[21,22,26] of about 25 meV is visible within the first 50 fs for positive $T$, which is consistent with previous observations.[1] Energy transfer between the excitonic states also leads to signal amplitude changes within a few tens of fs, which, however, overlap with a static GSB background and are therefore not clearly visible. As mentioned above, the subtraction procedure to obtain Δ1Q F-PP ($\omega_t$, $T$) will lead to reliable results only in case of a Markovian bath. For the dimer, separate 2DES experiments have not revealed strong system–bath correlations, justifying the theoretical description of these systems by time-independent Redfield theory.[6,69] On the other hand, the occurrence of a dynamic Stokes' shift in the 1Q F-PP spectrum of the dimer marks a deviation from Markovian behavior. In that case, the line shapes exhibit changes over $T$ and care must be taken when interpreting the difference spectra because the Stokes-shift-induced spectral asymmetry in the 1Q F-PP spectrum can introduce artificial features into Δ1Q F-PP ($\omega_t$, $T$), potentially distorting the extracted transients. Hence, due to the spectral drift of the lower-exciton feature during positive $T$, the respective feature in Δ1Q F-PP ($\omega_t$, $T$) must be integrated exactly in an interval that corresponds to that of the lower-exciton peak in the excitation spectrum (i.e., at $T < 0$) to obtain a reasonable transient. We obtain Δ1Q F-PP ($\omega_t$, $T$) according to Eq. (1), which is then further processed by integrating across the spectral region corresponding to the lower single-exciton state of the dimer (1.70 eV to 1.77 eV) and plotting it as a function of $T$ (Figure 7b, dots). The resulting Δ1Q F-PP ($T$) signal decays immediately from 0 fs and selectively displays the dynamics of energy transfer between the higher to the lower exciton state. This decay can be fitted by a monoexponential to retrieve energy-transfer dynamics of 21.1 ± 1.8 fs (Figure 7b, blue line). This value is consistent with previous PP, F-PP and F-2DES experiments.[1,6,51,70]

Note that all F-PP spectra feature a narrow coherent spike around $T = 0$,[1] as also observed in the simulations in Fig. 6. However, compared to coherently detected PP spectroscopy, this spike is much less pronounced, and its negative impact is diminished in view of our experimental implementation. This is largely because fluorescence detection suppresses non-resonant contributions. In addition, by using a pulse shaper, we ensure that all pulses are as short as possible, possess a flat spectral phase, and that the relative time-zero between pump and probe pulses is well defined, thereby minimizing artifacts arising from pulse overlap.

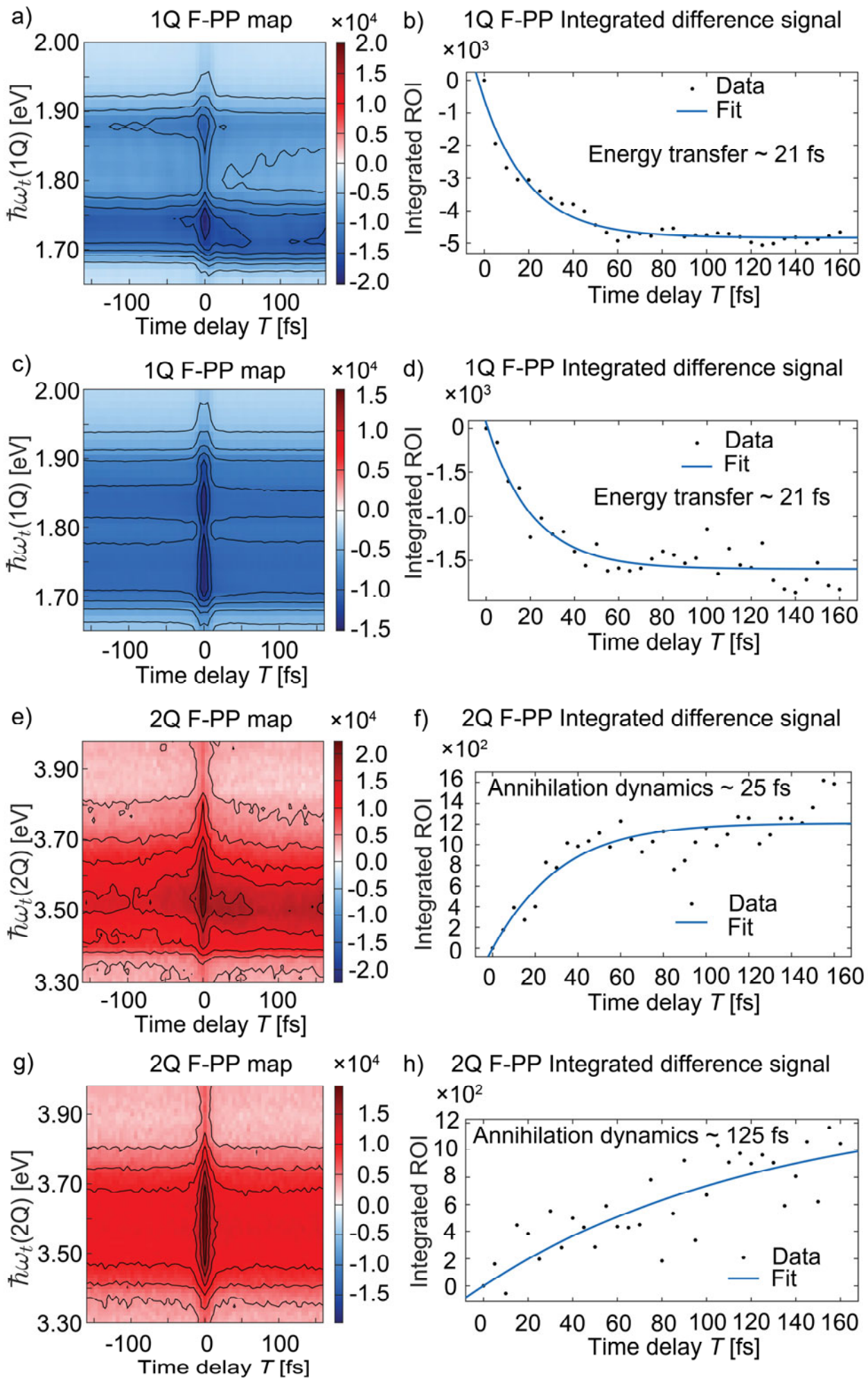

**Figure 7.** One-quantum (1Q) and two-quantum (2Q) F-PP spectroscopy experiment for a squaraine dimer and a hetero-polymer. (a) 1Q F-PP map of the squaraine heterodimer [SQA–SQB] in toluene. (b) Isolated population dynamics obtained by integrating Δ1Q F-PP ($\omega_t$, $T$) between 1.70 eV and 1.77 eV, displaying the energy-transfer dynamics between the single-exciton states, with a fit resulting in a time constant of 21.1 ± 1.8 fs. (c) 1Q F-PP map of the squaraine heteropolymer $[SQA–SQB]_5$ in toluene that consists of five dimeric units. (d) Kinetics of energy transfer obtained by integrating Δ1Q F-PP ($\omega_t$, $T$) between 1.70 eV and 1.77 eV, with a fit resulting in a time constant of 20.6 ± 3.2 fs. (e) 2Q F-PP map of the squaraine dimer. (f) Isolated population dynamics obtained by integrating Δ2Q F-PP ($\omega_t$, $T$) between 3.35 eV and 3.70 eV, energetically

corresponding to the doubly excited state. The Δ2Q F-PP ($T$) signal displays the dynamics of exciton diffusion and annihilation as the doubly excited manifold relaxes in a two-step process with a transfer from the biexciton state to a mixed |F⟩ state formed due to double excitation of same subsystem, followed by internal conversion to the coupled |E⟩ states. The fit time constant of the rise of the Δ2Q F-PP ($T$) signal is 24.8 ± 3.7 fs, i.e., displaying almost immediate annihilation considering the finite pulse duration. (g) 2Q F-PP map of the squaraine heteropolymer. (h) Integrated Δ2Q F-PP ($T$) signal of the polymer between 3.35 eV and 3.65 eV, with a fit time constant of 125.4 ± 65.6 fs. The rise is slower than in the dimer due to exciton diffusion along the polymer chain before annihilation.

In the case of the polymer (Figure 7c), we find a similarly fast energy-transfer rate of 20.6 ± 3.2 fs after analogous processing and fitting (Figure 7d). Such similar timescales make sense because both the polymer and the dimer are composed of the same SQA and SQB units between which energy transfer occurs. Comparing the 1Q F-PP spectrum of the dimer (Figure 7a) and the polymer (Figure 7c), we observe that the polymer shows almost no noticeable Stokes shift whereas the dimer does. This can be understood as a consequence of a large incoherent mixing contribution in the polymer, i.e., a dominating static GSB background due to the larger number of chromophores and a lower degree of exciton delocalization compared to the dimer.[66]

The sixth-order 2Q F-PP spectra are shown in Figure 7e for the dimer and in Figure 7g for the polymer. The annihilation kinetics were retrieved by spectrally integrating Δ2Q F-PP ($\omega_t$, $T$) and fitting according to Eq. (8). This separates the energy-transfer and annihilation components from the rise of the Δ2Q F-PP ($T$) signal. The value of $k_T$ was taken from the energy-transfer rate obtained from Δ1Q F-PP ($T$). This value is also close to the time constant determined in previous studies.[66] Note that the second term in Eq. (8), which is associated with the energy-transfer dynamics, depends on the difference of the transition dipole moments $\mu_{A0}$ and $\mu_{B0}$, i.e., from the ground state to the higher single-exciton state and from the ground state to the lower single-exciton state, respectively. The last term in Eq. (8) displays the annihilation dynamics. One would retrieve pure annihilation dynamics in case of $\mu_{A0} = \mu_{B0}$ because of a complete cancellation of the energy-transfer term. The ratio of the dipole moments $\mu_{A0}$ and $\mu_{B0}$ therefore controls the extent of the contribution of energy-transfer dynamics to the Δ2Q F-PP ($T$) signal. To determine the ratio of the transition dipole moments for the fit, we set $\mu_{A0}$ to 1 and scaled $\mu_{B0}$ using an estimated ratio of $\mu_{A0}$ and $\mu_{B0}$ as we describe below.

Note that Eq. (8) was derived under the assumption of the impulsive limit[6]. The effect of the actual laser spectral profile must still be considered when interpreting experimental results. Specifically, the intensity of the laser spectrum at the excitonic transition frequencies influences the measured signal. In experimental scenarios, laser pulses have finite duration and spectral bandwidth, and thus the system is not excited uniformly across all frequencies. Thus, in practical terms, the values of the quantities $\mu_{A0}$ and $\mu_{B0}$ in Eq. (8) should be understood as effective quantities: they represent the product of the absolute magnitude square of the transition dipole moments and the laser intensity at the corresponding transition frequency. This refinement connects the theoretical expression in Eq. (8) to measurable quantities and accounts for the finite bandwidth of real laser pulses. The quantities $\mu_{A0}^2$ and $\mu_{B0}^2$ were experimentally estimated by taking the maximum signal intensities of the higher and lower bands respectively, in the 1Q F-PP ($\omega_t$, $-T$) map, evaluated around –160 fs (to avoid pulse overlap). This can be done because the 1Q F-PP spectrum at $T \leq 0$ provides the excitation spectrum of the system under investigation[1] and the transition probability between two states is proportional to the product of the square of its transition dipole moment and the laser intensity at the corresponding transition frequency.[71] For the dimer, the maximum signal intensity of the excitation spectrum of the higher exciton state is ~0.65 times that of the maximum of the lower exciton state which is a result of the overlap of the laser spectrum with the linear absorption spectrum (see Fig. S1a in the SI). In case of the polymer, a different laser spectrum was used and its overlap with the absorption spectrum (Fig. S1b in the SI) caused the excitation spectrum to show two peaks with comparable maximum intensities, each associated with the respective excitonic state.

In the case of the dimer, spectrally integrating over Δ2Q F-PP ($\omega_t$, $T$) between 3.35 eV and 3.70 eV yields a transient that exhibits a fast rise with an annihilation time constant of 24.8 ± 3.7 fs (Figure 7f).[6] This is because the excitons in the heterodimer tend to be in close spatial proximity which causes almost immediate annihilation.[6,7,65,72] In the heteropolymer, on the other hand, the excitons generally have to diffuse before they meet each other and can annihilate. This circumstance results in a much slower rise of the Δ2Q F-PP ($T$) signal with diffusion-limited annihilation of 125.4 ± 65.6 fs (Figure 7h). Note that the polydispersity of the polymer[66] introduces a distribution of dynamical timescales which contributes to the comparably large uncertainty of the annihilation time, together with the fact that noise is further increased by performing a subtraction on a sixth-order nonlinear signal that is subject to a large extent of incoherent mixing, as further explained in the following.

The 2Q F-PP signal is also subject to incoherent mixing effects. The lower signal-to-noise ratio of the Δ2Q F-PP ($T$) signal of the polymer (Figure 7h) compared to the dimer (Figure 7f), obtained under identical experimental conditions, is due to incoherent mixing. In particular, the presence of a larger number of subsystems in the polymer results in a reduced contrast of excited-state dynamics in both 1Q and 2Q F-PP. This means that higher-order fluorescence-detected coherent spectroscopy becomes more challenging in systems with a larger number of chromophores or subsystems and in general requires a detector with a high dynamic range to separate excited-state kinetics with good signal-to-noise ratio. This is because in systems with large incoherent mixing, almost the entire signal range of the detector is used up by the contribution from static ground-state pathways, leaving a small dynamic range to record the desired excited-state dynamics. On the other hand, higher-order fluorescence-detected coherent spectroscopy and 2Q F-PP in particular hold promise due to their ability to eliminate parasitic pathways that otherwise plague the study of multi-exciton dynamics.

## 4. Conclusion

In conclusion, we introduced the new method of two-quantum fluorescence-detected pump–probe (2Q F-PP) spectroscopy. This method is particularly suitable as a tool to characterize the dynamics of the biexciton manifold of multichromophore systems. We described a pulse-shaper-based fully collinear shot-to-shot measurement of F-PP data using phase cycling for the simultaneous acquisition of 1Q and 2Q F-PP signals. The method makes efficient use of the measurement time because single- and multi-exciton dynamics are extracted from one dataset without requiring 2D frequency resolution. We provided a description of incoherent mixing in 1Q and 2Q F-PP, considering $N$ coupled subsystems. While this usually makes it challenging to extract excited-state dynamics due to the existence of a static background, we utilized a post-processing strategy that involves subtraction of F-PP data for negative pump–probe delay $T$ from those of positive $T$. This provides F-PP difference spectra that are free from incoherent mixing effects and that exclusively display energy transfer and biexciton dynamics. In the case of 2Q F-PP difference spectra, we also eliminate incorrectly time-ordered parasitic 2Q signals that originate during pump and probe overlap. These signals are otherwise a major artifact in (coherently detected) higher-order time-resolved experiments and introduce errors during analysis for the retrieval of multiparticle dynamics encoded in higher-order signals.

We quantified the dynamics of exciton energy transfer and diffusion-limited exciton–exciton annihilation for a squaraine heterodimer and a squaraine heteropolymer. While both systems exhibit similar rates of energy transfer between neighboring chromophores due to the identical geometry of the constituting heterodimer units, exciton annihilation takes a longer time in the polymer owing to the difference in length through which the exciton can diffuse.

As an outlook, 2Q F-PP could be extended to nonlinear time-resolved fluorescence microscopy experiments. This would provide additional spatial resolution and facilitate, for example, the discrimination of exciton dynamics and exciton diffusion in different spatial domains of materials. The 2Q F-PP method and subsequent calculation of a difference signal to extract doubly excited-state dynamics can also be transferred to other action-detected observables such as photocurrents, photoelectrons, or photoions.[73–76]

ASSOCIATED CONTENT

The data that support the findings of this paper are openly available in Zenodo at https://doi.org/10.5281/zenodo.17143841.

**Supporting Information**

Experimental methods, absorption spectra of squaraine polymer and dimer, simulation parameters for 1Q F-PP and 2Q F-PP for the exemplary model system.

AUTHOR INFORMATION

**Corresponding Author**

Tobias Brixner − *Institut für Physikalische und Theoretische Chemie and Institute for Sustainable Chemistry & Catalysis with Boron, Am Hubland, Universität Würzburg, 97074 Würzburg, Germany*; https://orcid.org/0000-0002-6529-704X; Email: tobias.brixner@uni-wuerzburg.de

**Author**

Ajay Jayachandran − Institut für Physikalische und Theoretische Chemie, Universität Würzburg, Am Hubland, 97074 Würzburg, Germany; https://orcid.org/0000-0002-5265-0502

Stefan Mueller − Institut für Physikalische und Theoretische Chemie, Universität Würzburg, Am Hubland, 97074 Würzburg, Germany; https://orcid.org/0009-0002-6384-9282

Christoph Lambert – Institut für Organische Chemie, Universität Würzburg, Am Hubland, 97074 Würzburg, Germany; https://orcid.org/0000-0002-9652-9165

**Notes**

The authors declare no competing financial interests.

ACKNOWLEDGMENTS

The authors acknowledge funding by the European Research Council (ERC) within Advanced Grant No. 101141366 (T.B.). The authors thank Arthur Turkin and Maximilian H. Schreck for the synthesis of the squaraine samples. We also thank Pavel Malý for discussions on F-PP.

# Supplementary Information:

# Background-free measurement of exciton–exciton annihilation by two-quantum fluorescence-detected pump–probe spectroscopy

Ajay Jayachandran[1], Stefan Mueller[1], Christoph Lambert[2], and Tobias Brixner[1,3,*]

[1]*Institut für Physikalische und Theoretische Chemie, Universität Würzburg, Am Hubland, 97074 Würzburg, Germany*

[2]*Institut für Organische Chemie, Universität Würzburg, Am Hubland, 97074 Würzburg, Germany*

[3]*Institute for Sustainable Chemistry & Catalysis with Boron (ICB), Universität Würzburg, Am Hubland, 97074 Würzburg, Germany*

## S1. Experimental methods

We used pulses generated by a Ti:sapphire regenerative amplifier (Spitfire Pro, Spectra-Physics, 800 nm, 35 fs) with a repetition rate of 1 kHz. The pulses were used to produce a white-light continuum using a fused-silica hollow-core fiber (Ultrafast Innovations GmbH) filled with a gas mixture of 60% argon and 40% neon at a total pressure of 1.2 bar. The input beam into the hollow-core fiber was spatially stabilized through a beam-stabilization system (Aligna, TEM Messtechnik GmbH). The pulse trains used for the fluorescence-detected pump–probe (F-PP) experiments were produced using an acousto-optic programmable dispersive filter (Dazzler, Fastlite). The pulses were compressed by using the Dazzler in combination with a grism compressor (Fastlite).

To acquire F-PP spectra, four-pulse sequences were generated shot-to-shot in a fully rotating frame of the laser center reference frequency. The delay $\tau$ between the pump pulses was set to zero, while the pump–probe delay $T$ and the coherence time $t$ were sampled sequentially using the pulse shaper. The $T$ delay was scanned from −160 fs to 160 fs in 5 fs delay steps and the $t$ delay was scanned between 0 fs to 120 fs in 6 fs delay steps. In addition to the delays between the pulses, the pulse phases were varied using 50-fold ($1 \times 2 \times 5 \times 5$) nested phase cycling [1,2]. In such a pulse sequence, the two-fold phase cycling of the second pulse (i.e., using phases 0 and $\pi$) effectively implements chopping. This is because the $\pi$ phase difference introduced in the second phase-cycling step causes destructive interference between the two pump pulses at zero delay, thereby creating a chopped pump pulse. In addition, the five-step phase cycling of the probe pulses separates both the 1Q F-PP and the 2Q F-PP signal as the five-step phase cycling resolves any ambiguity between the phase coefficients of the two signals.

The excitation beam was focused onto the sample inside a capillary flow cuvette with a square cross section of 250 µm × 250 µm. The sample solution was continuously pumped using a micro annular gear pump (mzr-2942-cy, HNP Mikrosysteme GmbH).

The [SQA–SQB] squaraine heterodimer and the $[SQA–SQB]_5$ copolymer samples were synthesized according to the literature [3, 4]. The samples were dissolved in toluene (spectroscopy grade, Sigma-Aldrich) and the fluorescence emitted from the sample was detected using an avalanche photodiode (A-Cube S500-3, Laser Components). This light was recollimated using a pair of microscope objectives (04OAS010, CVI Melles Griot), attenuated with neutral density filters (FS-3R, Newport) to match a linear detector response level, and guided via an optical fiber (QP400-2-SR, Ocean Optics) to the detector.

The pulses were initially compressed to a near flat phase using collinear SHG-FROG which was carried out by using the pulse shaper and a spectrometer (HR 4000, Ocean Optics). The dimer was excited using pulses with a duration of 18 fs (FWHM). The excitation energy was 158 nJ at maximum constructive interference of all excitation pulses. For the polymer, we used 20 fs pulses with an excitation energy of 175 nJ. The center reference wavelength of the laser was 680 nm.

## S2. Absorption spectra of squarine polymer and dimer

In Figure S1, we show the overlap of the laser spectra used in the individual experiments along the absorption spectra of the squaraine dimer [SQA–SQB] and the polymer $[SQA–SQB]_5$.

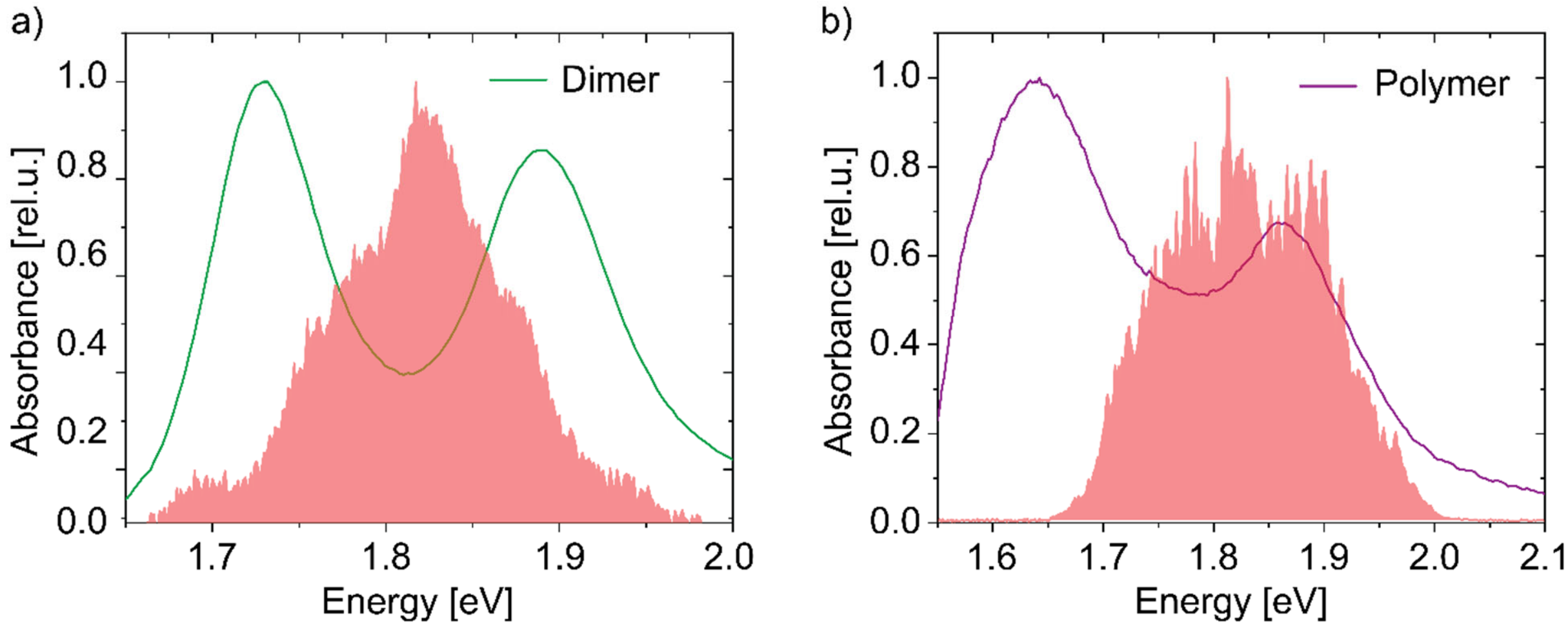

**Figure S1.** Absorption spectra of the [SQA–SQB] dimer and the $[SQA–SQB]_5$ polymer. **(**a) Overlap of the linear absorption spectrum of the squaraine dimer (green) with the laser spectrum (red shaded area). (b) Overlap of the linear absorption spectrum of the squaraine polymer (purple) and the laser spectrum (red shaded area). All spectra are normalized to their highest absolute values.

## S3. Simulation parameters

Simulations of the 1Q and 2Q F-PP spectra were carried out using the quantum-dynamics toolbox (QDT) [5] in Matlab R2024b. We summarize the parameters used in the simulation in Table S1. Briefly, we simulate the density matrix propagation in the Lindblad formalism [6] by explicitly including Gaussian pulses with finite duration. The model system is described by four energy levels in the exciton basis composed of a ground state $|0\rangle$, two single-exciton states $|A\rangle$ and $|B\rangle$, and a biexciton state $|2\rangle$. We incorporated distinct times for the relaxation within the single-exciton manifold, i.e., $|A\rangle \rightarrow |B\rangle$ (100 fs) and the relaxation of the biexciton state to the upper single-exciton states, i.e., $|2\rangle \rightarrow |A\rangle$ (30 fs).

**Table S1**. Parameters used for simulating 1Q and 2Q F-PP spectra.

| Quantity | | Value |
|---|---|---|
| Excitation parameters | Central energy $E_0$ | 1.79 eV |
| | Pulse duration (FWHM) | 10 fs |
| | External field amplitude | $0.3\times10^{-3}$ a.u. |
| Transition energies | $E_{B0}$ | 1.73 eV |
| | $E_{A0}$ | 1.88 eV |
| | $E_{20}$ | 3.61 eV |
| Transition dipole moments | $\mu_{B0}$ | 1.0 a.u. |
| | $\mu_{A0}$ | 0.9 a.u. |
| | $\mu_{2B}$ | 0.9 a.u. |
| | $\mu_{2A}$ | 1.0 a.u. |
| Population relaxation times | $T_{rel}$ (B → 0) | 1 ns |
| | $T_{rel}$ (A → B) | 100 fs |
| | $T_{rel}$ (2 → A) | 30 fs |
| Pure dephasing times | $T_{deph,B0}$ | 100 fs |
| | $T_{deph,A0}$ | 100 fs |
| | $T_{deph,BA}$ | 100 fs |
| | $T_{deph,2A}$ | 100 fs |
| | $T_{deph,2B}$ | 100 fs |
| | $T_{deph,20}$ | 50 fs |

To simulate the 1Q and 2Q F-PP signals, we first conducted a simulation with three pulses with individual phases $\varphi_1$, $\varphi_2$, and $\varphi_3$ using a 1×5×5 phase-cycling scheme in a fully rotating frame of the central energy of the pulses. In that sequence, the first pulse corresponds to the pump pulse whereas the last two pulses are the probe pulses. The time delay between the pump and the first probe pulse $T$ was sampled from –200 fs to 200 fs in 81 steps whereas the coherence time $t$ between the probe pulses was scanned from 0 to 84 fs in 15 steps. Based on these sampling parameters and the parameters listed in Tab. S1, we calculated the raw signal $S_{unchopped}(T, t, \varphi_{21}, \varphi_{31})$, where $\varphi_{21} = \varphi_2 - \varphi_1$ and $\varphi_{31} = \varphi_3 - \varphi_1$. We then repeated the simulation with the amplitude of the first pulse set to zero to simulate chopping of the pump pulse to obtain $S_{chopped}(T, t, \varphi_{21}, \varphi_{31})$. After subtracting $S_{chopped}(T, t, \varphi_{21}, \varphi_{31})$ from $S_{unchopped}(T, t, \varphi_{21}, \varphi_{31})$, we applied signal-specific phase weights to the resulting dataset $S(T, t, \varphi_{21}, \varphi_{31})$ according to

$$S_{1Q\ \text{F-PP}}(T,t) = \frac{1}{25}\sum_{b=0}^{4}\sum_{c=0}^{4} S(T,t,b\Delta\varphi_{21},c\Delta\varphi_{31})\exp(-ib\Delta\varphi_{21})\exp(ic\Delta\varphi_{31}), \quad \text{(S1)}$$

where $\varphi_{21} = b\Delta\varphi_{21}$, $\varphi_{31} = c\Delta\varphi_{31}$, with the phase increments $\Delta\varphi_{21} = \Delta\varphi_{31} = \frac{2\pi}{5}$, and

$$S_{2\text{Q F-PP}}(T,t) = \frac{1}{25}\sum_{b=0}^{4}\sum_{c=0}^{4} S(T,t,b\Delta\varphi_{21},c\Delta\varphi_{31}) \exp(-2ib\Delta\varphi_{21}) \exp(2ic\Delta\varphi_{31}), \quad \text{(S2)}$$

to extract the 1Q F-PP and the 2Q F-PP signals, respectively [2]. The corresponding F-PP spectra were then obtained via Fourier transformation of the signals with respect to *t*.